\documentclass[a4papet,10pt]{article}
\usepackage[italian, english]{babel}
\usepackage{amsfonts,amssymb,mathtools}
\usepackage{graphicx}
\usepackage{amsmath}
\usepackage[latin1]{inputenc} 
\usepackage{bbm}
\usepackage{pstricks}
\usepackage{psfrag}
\usepackage{appendix}
\usepackage{amsthm}
\usepackage{dsfont}
\usepackage{stackrel}
\usepackage{eufrak}
\usepackage{mathrsfs}
\usepackage{enumerate}
\usepackage{cite}
\usepackage{chngcntr}
\counterwithin{figure}{section}
\usepackage[a4paper,top=30mm,bottom=30mm,left=35mm,right=35mm]{geometry}
\usepackage{appendix}

\numberwithin{equation}{section}

\newcommand{\prob}{\mathbb{P}}

\newcommand{\Ex}{\mathbb{E}}

\newcommand{\paex}{\prob_o\mbox{-a.e.\ }x}
\newcommand{\Rl}{\mathbb{R}}

\newcommand{\X}{\{X_t\}_{t\ge 1}}

\newcommand{\bin}{\mathbb{N}_2}
\newcommand{\cov}{\mathrm{cov}}

\newtheoremstyle{mystyle}
  {}
  {}
  {\itshape}
  {}
  {\bfseries}
  {.}
  { }
  {\thmname{#1}\thmnumber{ #2}\thmnote{ (#3)}}

\theoremstyle{mystyle}
\newtheorem{theorem}{Theorem}[section]
\newtheorem{corollary}{Corollary}[section]
\newtheorem{lemma}{Lemma}[section]
\newtheorem{proposition}{Proposition}[section]

\title{Renewal model for dependent binary sequences}

\author{Marco Zamparo\footnote{Dipartimento di Fisica, Universit\`a degli Studi di Bari and INFN, Sezione di Bari,
    via Amendola 173, \phantom{aaz} 70126 Bari, Italy
    \newline \phantom{aaz} E-mail: \texttt{marco.zamparo@uniba.it}
\newline \phantom{aaz} ORCIDiD: 0000-0002-1336-0158}}

\date{}

\begin{document}  
\maketitle

\begin{abstract}
We suggest to construct infinite stochastic binary sequences by
associating one of the two symbols of the sequence with the renewal
times of an underlying renewal process. Focusing on stationary binary
sequences corresponding to delayed renewal processes, we investigate
correlations and the ability of the model to implement a prescribed
autocovariance structure, showing that a large variety of
subexponential decay of correlations can be accounted for. In
particular, robustness and efficiency of the method are tested by
generating binary sequences with polynomial and stretched-exponential
decay of correlations. Moreover, to justify the maximum entropy
principle for model selection, an asymptotic equipartition property
for typical sequences that naturally leads to the Shannon entropy of
the waiting time distribution is demonstrated.  To support the
comparison of the theory with data, a law of large numbers and a
central limit theorem are established for the time average of general
observables.\\


\end{abstract}

\section{Introduction}

Binary processes arise in natural and social sciences any time a
phenomenon is dichotomized according to the occurrence or
non-occurrence of a property of interest.  Modeling, statistical
inference from data, and generation of dependent binary sequences,
both finite and infinite, have thus received a lot of attention in a
number of disciplines, such as statistical physics \cite{Riccardo},
system biology \cite{Gene}, computational neuroscience
\cite{Neuroscience}, actuarial and financial sciences
\cite{ActuarialFinancialSciences}, and machine learning
\cite{DeepLearning}, among others. While finite sequences are
generally associated with graphical models \cite{Wainwright1},
infinite sequences are usually understood as time series.

Modeling and generating binary sequences with prescribed correlations
is one of the major problems to be faced \cite{Riccardo}. For finite
sequences, the solution to the problem is in principle represented by
the Ising model, which is an exponential family able to reproduce all
inputs in (the interior of) the correlation polytope
\cite{Wainwright1}.  For infinite sequences, the problem of
constructing binary variables with given correlations is more
challenging since there is no universal framework to refer to.
Indeed, finite-dimensional distributions structured according to the
Ising model are generally not consistent under marginalization, so
that they cannot be regarded as children of a common probability
measure associated with a stochastic process. On the other hand, the
Ising model is the only known parametric family that can cover all
correlations of a finite number of binary variables.  For this reason,
in order to describe and generate infinite binary sequences,
researchers have devised several approaches, each with its own
advantages and disadvantages, which can be grouped in autoregressive
models\cite{Regression,Markov1,Markov2,Markov3,Markov4,Markov5,Markov6,Markov7},
latent factor models \cite{Clipping1,Keenan,Iter1}, and mixed models
which combine autoregression with latent factors
\cite{Regression,Sung}.  Autoregressive models are Markov chains with
the property that the current probability of a symbol conditional on
the past history is determined by a linear function of the previous
outcomes \cite{Regression}, in a number corresponding to the order of
the chain, linearity being postulated to not have explosion of
parameters. These type of models can produce any exponential decay of
correlations, whereas subexponential decays are off limits due to
Markovianity and finite state space \cite{Bradley}. Latent factor
models rely on an underlying latent process, and the first proposal to
generate an infinite dependent binary sequence was to clip a latent
Gaussian process at a fixed level \cite{Clipping1}. Different latent
factor approaches are represented by mixture models whereby the binary
symbols are drawn independently with law determined by the realization
of an underlying process, such as a latent Gaussian process
\cite{Keenan} or a latent binary process \cite{Iter1}. Latent factor
approaches can introduce a long-range dependence between the binary
variables and can thus describe both exponential and subexponential
decays of correlations, such as polynomial decay. However, the
analysis of clipping performed in \cite{Signum} has revealed that
there are serious restrictions to the type of correlations that can be
produced. In fact, decay rates, in the case of exponential decay, and
decay exponents, in the case of polynomial decay, cannot be smaller
than a certain minimum threshold.  Similar restrictions have been
found in mixture models with latent binary inputs \cite{Iter1}. These
restrictions have been weakened in \cite{Iter2} by means of an
algorithm that by iterating the latent factor model of \cite{Iter1},
which transforms the latent binary input in a binary output, adds
dependence to an initial uncorrelated binary sequence step by
step. Other algorithms to generate dependent binary sequences and
their demonstration through numerical experiments are discussed in
\cite{Emrich,Beta1,Beta2,Algoritmi}.

In this paper we suggest and investigate a new latent factor approach
for infinite binary sequences that relies on renewal processes
\cite{Asmussen}. Our proposal is to associate one of the two symbols
of the sequence with the renewal times of a discrete-time delayed
renewal process, and the other symbol with all other times, thus
defining a regenerative phenomenon a la Kingman \cite{Kingman}.  Delay
makes it possible to construct stationary binary sequences. This way,
we can offer a flexible and powerful model that is under full
mathematical control and that is able to implement a large variety of
subexponential prescriptions for the correlations, from polynomial to
stretched-exponential decays as we shall demonstrate. Furthermore,
reflecting an underlying renewal structure, generation of the process
in numerical experiments is immediate.

The paper is organized as follows. In section \ref{sec:model} we
introduce the model and discuss its Markovian limit. Section
\ref{sec:correlations} is devoted to correlations. In this section we
prove the fundamental property of the model that binary symbols that
follows a renewal time are independent of previous symbols and we show
that the autocovariance solves a renewal equation.  Then we
investigate the direct problem of determining the asymptotics of
correlations from the waiting time distribution and the inverse
problem of associating a renewal structure to a prescribed
autocovariance for binary symbols. In section \ref{sec:empiricalmeans}
we prove limit theorems for the probability of typical sequences and
for the time averages of general observables. In particular, we
demonstrate an asymptotic equipartition property that naturally
introduces the Shannon entropy of the waiting time distribution and
leads to the maximum entropy principle for model selection
\cite{Jaynes1,Jaynes2}. Then we provide a law of large numbers and a
central limit theorem for the normal fluctuations of empirical means,
and we apply them to estimation of the waiting time distribution and
of the autocovariance from data. These results cannot be deduced from
the standard limit theory of regenerative processes with independent
cycles \cite{Asmussen} and require new arguments.  Conclusions are
drawn in section \ref{sec:conclusions}. The most technical
mathematical proofs are reported in the appendices in order to not
interrupt the flow of the presentation.

\section{Definition of the model}
\label{sec:model}

Let on a probability space $(\Omega,\mathscr{F},\prob)$ be given
positive integer-valued random variables $S_1,S_2,\ldots$. We
construct a binary stochastic sequence $X:=\X$ with entries valued in
$\bin:=\{0,1\}$ by supposing that the variable $S_n$ is the
\textit{waiting time} for the $n$th symbol 1 of the sequence, which
occurs at the \textit{renewal time} $T_n:=S_1+\cdots+S_n$. Thus,
$X_t:=1$ if $t\in\{T_n\}_{n\ge 1}$ and $X_t:=0$ otherwise for each
$t\ge 1$.  Our fundamental assumptions are $(i)$ that waiting times
are independent and $(ii)$ that $S_2,S_3,\ldots$ have the same
distribution, which could differ from the distribution of $S_1$. Under
these assumptions, the sequence $\{T_n\}_{n\ge 1}$ is a
\textit{renewal process}, \textit{delayed} if $S_1$ is not distributed
as $S_2$. We refer to \cite{Asmussen} for basics of renewal
theory. The reason for letting $S_1$ behave differently from the other
waiting times is that its distribution can be chosen in such a way
that the sequence $X$ is stationary, as explained by the following
theorem which is proved in Appendix \ref{proof:stationarity}. We
recall that the process $X$ is \textit{stationary} if
$\{X_{t+1}\}_{t\ge 1}$ is distributed as $\X$, where $\X$ is
understood as a random element in the set $\bin^\infty$ of all binary
sequences equipped with the cylinder $\sigma$-field $\mathscr{B}$.
Hereafter, $\Ex$ denotes expectation with respect to $\prob$.
\begin{theorem}
  \label{th:stationarity}
  The binary sequence $X$ is stationary if and only if
  $\Ex[S_2]<\infty$ and for every $s\ge 1$
  \begin{equation*}
\Ex[S_2]\,\prob[S_1=s]=\prob[S_2\ge s].
\end{equation*}
\end{theorem}

Theorem \ref{th:stationarity} recovers well-known conditions for the
delayed renewal chain $\{T_n\}_{n\ge 1}$ to be stationary, that means
that the statistical properties of the number of renewals in a given
temporal window are invariant with respect to time shifts
\cite{Asmussen}. Under the hypothesis of stationarity, which is now
made, the only input of the model $X$ is the distribution of the
waiting time $S_2$. Depending on the need, we refer to its density
$p:=\prob[S_2=\cdot\,]$ or to its tail $Q:=\prob[S_2>\cdot\,]$. Thus,
according to Theorem \ref{th:stationarity}, we suppose that
$\mu:=\Ex[S_2]=\sum_{s\ge 1}sp(s)=\sum_{t\ge 0}Q(t)<\infty$ and that
$\prob[S_1=s]=(1/\mu)\sum_{n\ge s}p(n)$ for every $s\ge 1$. We notice
that $\Ex[X_1]=\prob[T_1=1]=\prob[S_1=1]=1/\mu$.

The analysis of the finite-dimensional distributions reveals that if
the sequence $X$ is stationary, then it is \textit{time-reversible} in
the sense that $(X_t,\ldots,X_2,X_1)$ is distributed as
$(X_1,X_2,\ldots,X_t)$ for all $t$. The finite-dimensional
distributions of $X$ are provided by the next proposition, which is
demonstrated in Appendix \ref{proof:dist}. We make the usual
conventions that an empty sum is assumed to be 0 and an empty product
is assumed to be 1.
\begin{proposition}
\label{prop:dist}
For any integer $t\ge 1$ and numbers $x_1,x_2,\ldots,x_t$ in $\bin$
\begin{align}
  \nonumber
  \pi_t(x_1,x_2,\ldots,x_t):&=\prob\big[X_1=x_1,X_2=x_2,\ldots,X_t=x_t\big]\\
  \nonumber
  &=\frac{1}{\mu}\bigg[\sum_{s\ge t}Q(s)\bigg]^{\prod_{n=1}^t(1-x_n)}
  \prod_{i=1}^t\big[Q(i-1)\big]^{\prod_{n=1}^{i-1}(1-x_n)x_i}\\
  \nonumber
  &\,\cdot\,\,\prod_{i=1}^{t-1}\prod_{j=i+1}^t\big[p(j-i)\big]^{x_i\prod_{n=i+1}^{j-1}(1-x_n)x_j}\prod_{j=1}^t\big[Q(t-j)\big]^{x_j\prod_{n=j+1}^t(1-x_n)}\\
  \nonumber
  &=\pi_t(x_t,\ldots,x_2,x_1)
\end{align}
with $0^0:=1$.
\end{proposition}

Proposition \ref{prop:dist} can be used to compute the conditional
probability of a symbol given the past and allows us to get some
insight into the dependence structure of the stochastic sequence
$X$. Let us denote by $l_t(x_1,\ldots,x_t)$ the position of the first
symbol $1$ in the binary string $(x_1,\ldots,x_t)$:
$l_t(x_1,\ldots,x_t):=l$ if $x_l=1$ and $x_1=\cdots=x_{l-1}=0$ for
some $l\le t$, and $l_t(x_1,\ldots,x_t):=\infty$ if
$x_1=\cdots=x_t=0$.  Proposition \ref{prop:dist} shows through easy
manipulations that for any positive integer $t$ and binary numbers
$x_1,\ldots,x_t$
\begin{equation}
  \prob\big[X_{t+1}=1\big|X_t=x_1,\ldots,X_1=x_t\big]=
  \begin{cases}
    \displaystyle{\frac{Q(t)}{\sum_{s\ge t}Q(s)}} & \mbox{if }l_t(x_1,\ldots,x_t)=\infty;\\[1em]
    \displaystyle{\frac{p(l)}{Q(l-1)}} & \mbox{if }l_t(x_1,\ldots,x_t)=:l<\infty
  \end{cases}
  \label{cond_prob}
\end{equation}
provided that $\prob[X_1=x_t,\ldots,X_t=x_1]>0$. Pay attention to the
inverse order of $x_1,\ldots,x_t$ in the conditioning event. Formula
(\ref{cond_prob}) draws a link between our model and the so-called
``stochastic chains with memory of variable length'' \cite{SCMVL}. In
this class of models, which can be built on a larger alphabet than
$\bin$, the number of variables involved in the conditioning event is
determined by a ``context length function'', which itself depends on
the past variables. The context length function of our model is
exactly the function $l_t$ that maps any binary string
$(x_1,\ldots,x_t)$ in $l_t(x_1,\ldots,x_t)$. Since the function $l_t$
is unbounded as the time $t$ goes on, our model constitutes a
stochastic chain having in general memory of unbounded variable
length.

In spite of an unbounded context length, the process $X$ is a Markov
chain for certain particular waiting time distributions $p$.  The
sequence $X$ is a \textit{Markov chain} of order $M\ge 1$ if the
conditional probability of a symbol given the past depends only on the
state of the last $M$ variables. We can include sequences of
i.i.d.\ random variables in this definition by letting $M$ to take the
value zero.  The requirement that the conditional probability
(\ref{cond_prob}) is independent of $x_{M+1},\ldots,x_t$ for
Markovianity of order $M\ge 0$ results in the following corollary of
Proposition \ref{prop:dist}, whose proof is reported in Appendix
\ref{proof:Markovchain}.
\begin{corollary}
  \label{Markovchain}
  The binary sequence $X$ is a Markov chain of order $M\ge 0$ if and only
  if there exists a real number $\lambda\in[0,1)$ such that
  $p(M+s+1)=\lambda^sp(M+1)$ for all $s\ge 1$.
\end{corollary}

\section{Correlations}
\label{sec:correlations}

The binary sequence $X:=\X$ is a \textit{regenerative process} with
regeneration points $\{T_n\}_{n\ge 1}$ and independent cycles
\cite{Asmussen}. This means that the past and the future of each
renewal event, i.e.\ of each symbol 1, are independent. Such
conditional independence is stated by the following proposition, whose
proof is provided in Appendix \ref{proof:condind}.
\begin{proposition}
\label{prop:condind}
For every positive integers $s\le t$ and binary numbers
$x_1,\ldots,x_t$
\begin{align}
\nonumber
  \prob\big[X_1=x_1,\ldots,X_t=x_t\big|X_s=1\big]=\,\,&\prob\big[X_1=x_1,\ldots,X_s=x_s\big|X_s=1\big]\\
  \nonumber
  \cdot\,\,\,&\prob\big[X_s=x_s,\ldots,X_t=x_t\big|X_s=1\big].
\end{align}
\end{proposition}


Conditional independence associated with renewals is exploited in this
section to study correlations.  The \textit{covariance} of two random
variables $Y$ and $Z$ on the probability space
$(\Omega,\mathscr{F},\prob)$ is denoted by $\cov[Y,Z]$:
\begin{equation*}
  \cov[Y,Z]:=\Ex\big[(Y-\Ex[Y])(Z-\Ex[Z])\big]=\Ex[YZ]-\Ex[Y]\,\Ex[Z].
\end{equation*}
For every $t\ge 0$, the \textit{autocovariance}
$\rho_t:=\cov[X_1,X_{t+1}]$ reads $\rho_t=c_t-c_0^2$ with
\begin{equation*}
c_t:=\Ex[X_1X_{t+1}].
\end{equation*}
Notice that $c_0=\Ex[X_1]=1/\mu$.  The next proposition, which is
based on Proposition \ref{prop:condind} and is proved in Appendix
\ref{proof:c}, shows that the sequence $c:=\{c_t\}_{t\ge 0}$ solves a
renewal equation with waiting time distribution $p$. As a consequence,
the renewal theorem \cite{Erdos} gives
$\lim_{t\uparrow\infty}c_t=c_0/\mu=c_0^2$, namely
$\lim_{t\uparrow\infty}\rho_t=0$, provided that $p$ is aperiodic. The
probability distribution $p$ is \textit{aperiodic} if there is no
proper sublattice of $\{1,2,\dots\}$ containing its support.
\begin{proposition}
\label{prop:c}
For every integer $t\ge 1$
\begin{equation*}
 c_t=\sum_{s=1}^tp(s)\,c_{t-s}.
  \end{equation*}
\end{proposition}

By resorting to the literature on the renewal equation, we investigate
the direct problem of determining the asymptotics of the sequence $c$,
and hence of the autocovariance, from a given waiting time
distribution $p$, as well as the inverse problem of finding conditions
on a prescribed $c$ such that $c$ solves a renewal equation with some
probability distribution $p$ of finite mean. The latter aims to answer
the question of which autocovariance structures can be reproduced by
our model.  Before that, we want to point out that the autocovariance
of the binary sequence $X$ describes the time dependence of any
temporal correlations, as shown by the following technical lemma that
explores temporal correlations of general observables. The lemma is
proved in Appendix \ref{proof:covYZ} and will be used in section
\ref{sec:empiricalmeans} to address the mixing properties of $X$.
\begin{lemma}
\label{lem:covYZ}
Fix positive integers $m$ and $n$ and let $f$ and $g$ be two real
functions on $\bin^m$ and $\bin^n$, respectively, such that
$f(0,\ldots,0)=g(0,\ldots,0)=0$. Set $Z_t:=g(X_t,\ldots,X_{t+n-1})$
for $t\ge 1$. Then, for all $t\ge 1$
\begin{align}
\nonumber
\cov\big[f(X_1,\ldots,X_m),Z_{m+t}\big]&=\sum_{i=1}^m\sum_{j=1}^n C_{i,j}(t)\,
\frac{\Ex[\mathds{1}_{\{S_1=i\}}f(X_m,\ldots,X_1)]}{\prob[S_1=i]}\,\frac{\Ex[\mathds{1}_{\{S_1=j\}}Z_1]}{\prob[S_1=j]},
\end{align}
where for $i\ge 1$ and $j\ge 1$ 
\begin{equation*}
  C_{i,j}(t):=\sum_{u=1}^i\sum_{v=1}^j p(i-u)\,\rho_{u+t+v-2}\,p(j-v)
\end{equation*}
with $p(0):=-1$.
\end{lemma}

\subsection{Autocorrelation: direct problem}

Let us study the asymptotics of the autocovariance for a given waiting
time distribution.  Pure exponential decay of correlations can be
described by Markov chains as identified by Corollary
\ref{Markovchain} and is somehow trivial. We shall touch on the
exponential behavior of the autocovariance when dealing with the
inverse problem. Here we focus on subexponential decays that account
for long-range dependence that cannot be explained by Markov
processes. A natural setting for subexponentiality in renewal theory
was given in \cite{Chover}. Let the symbol $\sim$ denotes asymptotic
equivalence for sequences: $a_t\sim b_t$ means
$\lim_{t\uparrow\infty}a_t/b_t=1$. Following \cite{Chover}, we say
that a positive real sequence $a:=\{a_t\}_{t\ge 0}$ belongs to the
class $\mathscr{S}$ of \textit{subexponential sequences} if
$A:=\sum_{t\ge 0}a_t<\infty$, $a_{t+1}\sim a_t$, and
$\sum_{n=0}^ta_na_{t-n}\sim 2Aa_t$. The requirement $a_{t+1}\sim a_t$
prevents $a$ from growing exponentially, thus justifying the
terminology ``subexponential''. The asymptotic behavior of the
autocovariance $\rho_t$ can be characterized in general when there
exists $\lambda\in(0,1]$ such that $\{\lambda^{-t}Q(t)\}_{t\ge
    0}\in\mathscr{S}$, namely when the tail probability $Q$ has
  exponential rate $\lambda$ and a subexponential correction that
  forms a sequence in $\mathscr{S}$. The case $\lambda=1$ corresponds
  to a pure subexponential behavior.  In fact, Theorem 3.2 of
  \cite{Embrechts} for the rate of convergence of renewal sequences
  gives the following result.
\begin{theorem}
\label{th:subexp}
Assume that $p$ is aperiodic and that $\{\lambda^{-t} Q(t)\}_{t\ge
  0}\in\mathscr{S}$ for some $\lambda\in(0,1]$. Then
\begin{equation*}
\rho_t\sim\frac{\sum_{n>t}Q(n)}{\mu\big[\sum_{n\ge 0}\lambda^{-n} Q(n)\big]^2}.
\end{equation*}
\end{theorem}

Subexponential behaviors that find wide application are polynomial
decays, which fall under the umbrella of regular variation, and
Weibull-type decays represented by stretched exponentials. We now
suppose $\lambda=1$ and discuss these decays in some detail. We stress
that the necessary condition $\sum_{t\ge 0}Q(t)<\infty$ for the
sequence $\{Q(t)\}_{t\ge 0}$ to belong to $\mathscr{S}$ is satisfied
since $\sum_{t\ge 0}Q(t)=\mu<\infty$ by the hypothesis of
stationarity.

\subsubsection{Polynomial decay}

A positive sequence $a:=\{a_t\}_{t\ge 0}$ is \textit{regularly
  varying} if there exists an index $\alpha\in\Rl$ and a slowly
varying function $\ell$ such that $a_t\sim t^\alpha\ell(t)$.  A real
measurable function $\ell$ is \textit{slowly varying} if it is
positive on a neighborhood of infinity, say $(\tau,\infty)$ with some
$\tau>0$, and satisfies the scale-invariance property
$\lim_{z\uparrow\infty}\ell(\eta z)/\ell(z)=1$ for any number
$\eta>0$. Trivially, a measurable function with a finite positive
limit at infinity is slowly varying. The simplest non-trivial example
is represented by the logarithm. We refer to \cite{RV} for the theory
of slow and regular variation. We stress that a slowly varying
function $\ell$ is dominated by polynomials in the sense that
$\lim_{z\uparrow\infty}z^\gamma\ell(z)=\infty$ and
$\lim_{z\uparrow\infty}z^{-\gamma}\ell(z)=0$ for all $\gamma>0$
according to Proposition 1.3.6 of \cite{RV}.  The uniform convergence
theorem \cite{RV} states that the scale-invariance property of slowly
varying functions actually is uniform for $\eta$ that belongs to each
compact set in $(0,\infty)$. This fact can be used to show that
$a_{t+1}\sim a_t$ if $a$ is a regularly varying sequence. Combined
with the dominated convergence theorem, it also shows that
$\sum_{n=0}^ta_na_{t-n}\sim 2\sum_{n=0}^{\lfloor
  t/2\rfloor}a_na_{t-n}\sim 2Aa_t$ when $A:=\sum_{t\ge 0}a_t<\infty$,
$\lfloor t/2\rfloor$ denoting the integer part of $t/2$. Thus, any
summable regularly varying sequence is an element of
$\mathscr{S}$. Summability imposes the restriction $\alpha\le -1$ to
the index.

These arguments show that if $Q(t)\sim t^{-\gamma-1}\ell(t)$ with an
exponent $\gamma>0$ and an arbitrary slowly varying function $\ell$,
then $\{Q(t)\}_{t\ge 0}\in\mathscr{S}$. In such case we have the
asymptotic equivalence $\sum_{n>t}Q(n)\sim
(1/\gamma)t^{-\gamma}\ell(t)$ by Proposition 1.5.10 of \cite{RV}.
Thus, we get the following corollary of Theorem \ref{th:subexp}.
\begin{corollary}
  \label{cor:subexp1}
Assume that $p$ is aperiodic and $Q(t)\sim t^{-\gamma-1}\ell(t)$ with
an exponent $\gamma>0$ and a slowly varying function $\ell$. Then
\begin{equation*}
\rho_t\sim \frac{\ell(t)}{\gamma\mu^3t^{\gamma}}.
\end{equation*}
\end{corollary}

In contrast to the latent factor models analyzed in \cite{Iter1} and
\cite{Signum}, which can account for polynomial decay of correlations
but not for too small exponents, Corollary \ref{cor:subexp1} of
Theorem \ref{th:subexp} shows that a renewal structure is able to
describe polynomial decays of the autocovariance with any exponent
$\gamma>0$. Actually, the hypothesis $\gamma>0$ is not necessary and
we can also have $\gamma=0$, so that the autocovariance decays slower
than any polynomial, but in this case the asymptotic behavior of
$\{\sum_{n>t}Q(n)\}_{t\ge 0}$ cannot be resolved in general. Notice
that summability of $\{Q(t)\}_{t\ge 0}$ when $\gamma=0$ imposes
restrictions on $\ell$.  For instance, if $Q(t)\sim t^{-1}(\ln
t)^{-\beta-1}$ with a number $\beta>0$, then $\{Q(t)\}_{t\ge
  0}\in\mathscr{S}$ and Theorem \ref{th:subexp} gives
$\rho_t\sim(1/\mu^3)\sum_{n>t}Q(n)\sim(1/\beta\mu^3)(\ln t)^{-\beta}$.

\subsubsection{Stretched-exponential decay}

In \cite{Chover} the following sufficient condition for
subexponentiality was proposed. Let $h$ be a continuously
differentiable real function on a neighborhood of infinity, say
$(\tau,\infty)$ with some $\tau>0$, such that its derivative $h'$
enjoys the properties that $-z^2h'(z)$ is increasing to infinity with
respect to $z$ and that $\int_\tau^\infty
e^{\frac{1}{2}z^2h'(z)}dz<\infty$. Then, a positive sequence
$a:=\{a_t\}_{t\ge 0}$ such that $A:=\sum_{t\ge 0}a_t<\infty$ and
$a_{t+1}\sim a_t\sim e^{-th(t)}$ satisfies $\sum_{n=0}^ta_na_{t-n}\sim
2Aa_t$, and hence belongs to $\mathscr{S}$. We have used this
criterion to investigate the asymptotic behavior of the autocovariance
when $Q(t)\sim e^{-t^\beta\ell(t)}$ with a stretching exponent
$\beta\in(0,1)$ and some function $\ell$. The following corollary of
Theorem \ref{th:subexp}, which is proved in Appendix
\ref{proof:stretched}, gives sufficient conditions on $\ell$ that
imply $\{Q(t)\}_{t\ge 0}\in\mathscr{S}$. We point out that $\ell$ is
slowly varying under those conditions.
\begin{corollary}
\label{corol:stretched}
Assume that $p$ is aperiodic and $Q(t)\sim e^{-t^\beta\ell(t)}$ with a
stretching exponent $\beta\in(0,1)$ and a twice continuously
differentiable positive function $\ell$ on a neighborhood of infinity
that satisfies $\lim_{z\uparrow\infty}z\ell'(z)/\ell(z)=0$ and
$\lim_{z\uparrow\infty}z^2\ell''(z)/\ell(z)$ exists.  Then
\begin{equation*}
\rho_t\sim\frac{t^{1-\beta}}{\mu^3\beta\ell(t)} e^{-t^\beta\ell(t)}.
\end{equation*}
\end{corollary}

To conclude, we observe that the hypothesis $\beta<1$ can be relaxed
in favor of $\beta=1$ to come fairly close to exponential decay of
correlations while staying in the framework of subexponential
sequences. If for example $Q(t)\sim e^{-t(\ln t)^{-\gamma}}$ with some
number $\gamma>0$, then $Q(t+1)\sim Q(t)\sim e^{-th(t)}$ with
$h(z):=(\ln z)^{-\gamma}$. This function $h$ satisfies the above
sufficient condition for subexponentiality, so that
$\rho_t\sim(1/\mu^3)\sum_{n>t}Q(n)\sim (1/\mu^3)(\ln t)^\gamma
e^{-t(\ln t)^{-\gamma}}$ by Theorem \ref{th:subexp}.

\subsection{Autocorrelation: inverse problem}

Let us now investigate the possibility to implement a prescribed
autocovariance.  Here the focus is on short time scales since Theorem
\ref{th:subexp} and its Corollaries \ref{cor:subexp1} and
\ref{corol:stretched} already settle the issue on long time scales,
demonstrating that a large class of asymptotic prescriptions can be
obtained.  We want to understand conditions on a non-negative sequence
$c:=\{c_t\}_{t\ge 0}$ under which there exists a waiting time
distribution $p$ of finite mean whose associated stationary binary
sequence $X:=\X$ satisfies $c_t=\Ex[X_1X_{t+1}]$ for all $t\ge 0$. For
simplicity, we assume that $c_t>0$ for every $t\ge 0$. In the light of
Theorem \ref{th:stationarity} and Proposition \ref{prop:c}, this is
tantamount to ask under which conditions on $c$ there exists a
probability distribution $p$ with the properties $\sum_{s\ge
  1}sp(s)=1/c_0$ and $\sum_{s=1}^tp(s)\,c_{t-s}=c_t$ for all $t\ge
1$. Such a $p$, if any, is uniquely defined by the renewal equation
and meets the requirement $\sum_{s\ge 1}sp(s)=1/c_0$ if and only if
$\lim_{t\uparrow\infty}c_t=c_0^2$. In fact, if $p$ is a probability
distribution such that $\sum_{s=1}^tp(s)\,c_{t-s}=c_t$ for all $t\ge
1$, then it is aperiodic since $p(1)=c_1/c_0>0$. This way, the renewal
theorem \cite{Erdos} gives $\lim_{t\uparrow\infty}c_t=c_0/\sum_{s\ge
  1}sp(s)$ if $\sum_{s\ge 1}sp(s)<\infty$ and
$\lim_{t\uparrow\infty}c_t=0$ if $\sum_{s\ge 1}sp(s)=\infty$.

Finding the minimal conditions on the sequence $c$ for the existence
of an associated waiting time distribution $p$ is a difficult task.
We stress that the problem consists in determining whether or not the
function $p$ that solves the problem $\sum_{s=1}^tp(s)\,c_{t-s}=c_t$
for every $t\ge 1$ is non-negative and sums to 1. However, there is a
sufficient condition that covers many applications. A sequence
$c:=\{c_t\}_{t\ge 0}$ is a \textit{Kaluza sequence} \cite{Kingman} if
$c_t>0$ and $c_{t-1}c_{t+1}\ge c_t^2$ for all $t\ge 1$. It follows
that $c_0>0$. The following theorem states that the hypothesis that
$c$ is a Kaluza sequence such that $\lim_{t\uparrow\infty}c_t=c_0^2$
guarantees the existence of an associated waiting time distribution of
finite mean.  The proof is provided in Appendix
\ref{proof:inverseproblem}.
\begin{theorem}
\label{th:inverseproblem}
Let $c:=\{c_t\}_{t\ge 0}$ be a Kaluza sequence such that
$\lim_{t\uparrow\infty}c_t=c_0^2$. Then, there exists a unique waiting
time distribution $p$ with the properties $\sum_{s\ge 1}sp(s)=1/c_0$
and $\sum_{s=1}^tp(s)\,c_{t-s}=c_t$ for all $t\ge 1$. As a
consequence, the stationary binary sequence $X:=\X$ associated with
$p$ satisfies $c_t=\Ex[X_1X_{t+1}]$ for all $t\ge 0$.
\end{theorem}

We point out that Theorem 3.2 of \cite{Embrechts} offers an inverse of
Theorem \ref{th:subexp}: if $c:=\{c_t\}_{t\ge 0}$ is a Kaluza sequence
such that $\lim_{t\uparrow\infty}c_t=c_0^2$ and $\{c_{t+1}-c_t\}_{t\ge
  0}\in\mathscr{S}$, then there exists a unique associated waiting
time distribution $p$ whose tail $Q$ enjoys the property
$(1/\mu^3)\sum_{n>t}Q(n)\sim \rho_t:=c_t-c_0^2$.

A practical criterion to recognize a Kaluza sequence that puts the
emphasis on the autocovariance is the following. Consider a sequence
$c:=\{c_t\}_{t\ge 0}$ defined by $c_0:=\xi$ and $c_t:=\xi^2+m
e^{-\phi(t)}$ for $t\ge 1$ with constants $\xi\in(0,1]$ and
  $m\in[0,\xi(1-\xi)]$ and a concave function $\phi$ such that
  $\phi(0)=0$ and $\lim_{z\uparrow\infty}\phi(z)=\infty$. Then, $c$ is
  a Kaluza sequence and $\lim_{t\uparrow\infty}c_t=c_0^2$. Indeed,
  $c_0=\xi^2+\xi(1-\xi)\ge \xi^2+me^{-\phi(0)}$, so that for each
  $t\ge 1$
\begin{align}
  \nonumber
  c_{t-1}c_{t+1}-c_t^2 &\ge[\xi^2+m e^{-\phi(t-1)}][\xi^2+m e^{-\phi(t+1)}]-[\xi^2+m e^{-\phi(t)}]^2\\
  \nonumber
  &=m\xi^2[e^{-\phi(t-1)}+e^{-\phi(t+1)}-2e^{-\phi(t)}]\\
  \nonumber
  &+m^2e^{-2\phi(t)}[e^{2\phi(t)-\phi(t-1)-\phi(t+1)}-1].
  \end{align}
This way, the concavity of $\phi$ and the consequent convexity of
$e^{-\phi}$ give $c_{t-1}c_{t+1}-c_t^2\ge 0$ for all $t\ge 1$. We have
thus proved the following corollary of Theorem
\ref{th:inverseproblem}.
\begin{corollary}
  \label{corollary:inverseproblem}
  Let $\xi\in(0,1]$ and $m\in[0,\xi(1-\xi)]$ be two constants and let
    $\phi$ be a concave function such that $\phi(0)=0$ and
    $\lim_{z\uparrow\infty}\phi(z)=\infty$. Then, there exists a
    unique waiting time distribution $p$ of finite mean whose
    associated stationary binary sequence $X:=\X$ satisfies
    $\Ex[X_1]=\xi$ and $\rho_t:=\cov[X_1,X_{t+1}]=m e^{-\phi(t)}$ for
    all $t\ge 1$.
\end{corollary}

\begin{figure}
  \centering
\includegraphics[height=5cm,width=10cm]{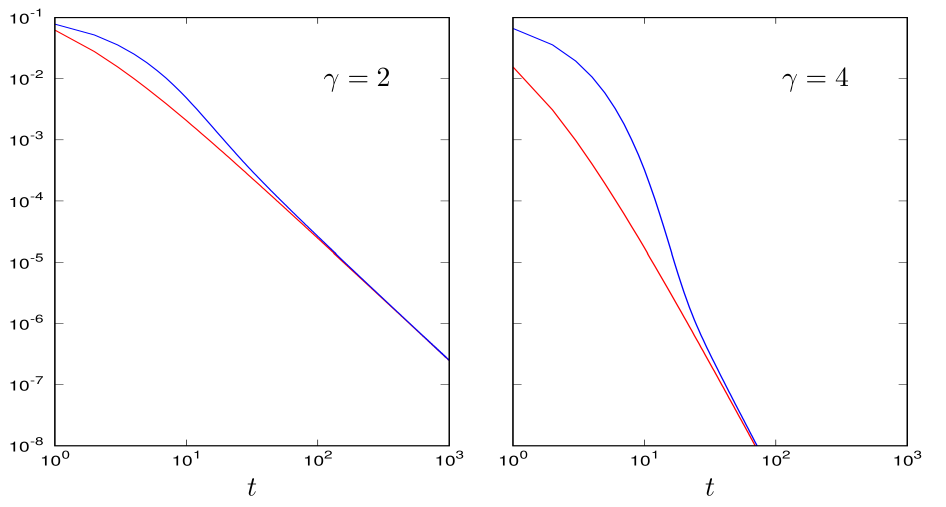}
\caption{Autocovariance $\rho_t$ (red) and $(1/\mu^3)\sum_{n>t}Q(n)$
  (blue) versus $t$ for the inverse problem with
  $c_t=1/4+(1/4)(1+t)^{-\gamma}$ for $t\ge 0$ with $\gamma=2$
  (left) and $\gamma=4$ (right).}
\label{fig:inverse_pol}
\vspace{0.8cm}
\includegraphics[height=5cm,width=10cm]{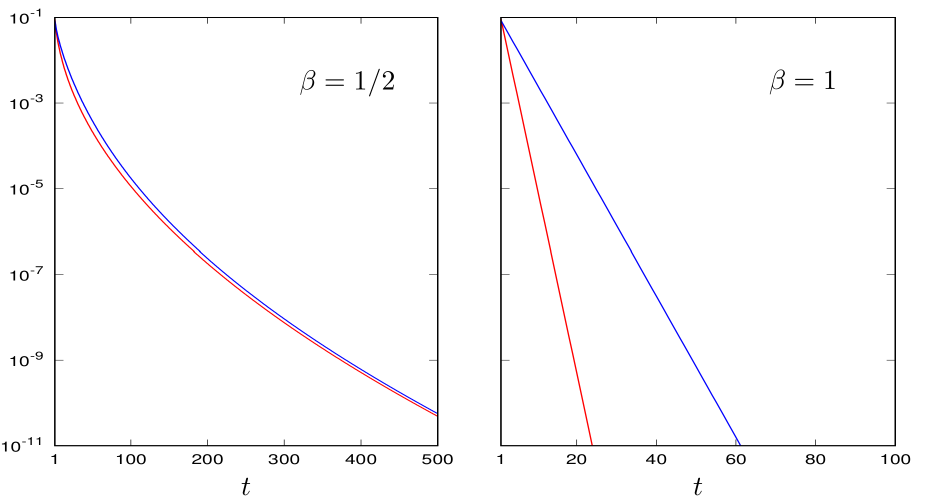}
\caption{Autocovariance $\rho_t$ (red) and $(1/\mu^3)\sum_{n>t}Q(n)$
  (blue) versus $t$ for the inverse problem with
  $c_t=1/4+(1/4)e^{-t^\beta}$ for $t\ge 0$ with $\beta=1/2$
  (left) and $\beta=1$ (right).}
\label{fig:inverse_str}
\end{figure}

We appeal to Corollary \ref{corollary:inverseproblem} to draw some
examples. Our model can implement the autocovariances
$\rho_t=m(1+t)^{-\gamma}$ and $\rho_t=me^{-\kappa t^\beta}$ for $t\ge
1$ with any real numbers $\xi\in(0,1]$, $m\in[0,\xi(1-\xi)]$,
  $\gamma>0$, $\kappa>0$, and $\beta\in(0,1]$. Figure
    \ref{fig:inverse_pol} compares $\rho_t$ with
    $(1/\mu^3)\sum_{n>t}Q(n)$ for the two inverse problems
    corresponding to the polynomial correlation structures
    $c_t=1/4+(1/4)(1+t)^{-\gamma}$ for $t\ge 0$ with exponents
    $\gamma=2$ and $\gamma=4$, respectively. We see that the earlier
    mentioned asymptotic equivalence
    $\rho_t\sim(1/\mu^3)\sum_{n>t}Q(n)$ is confirmed. Figure
    \ref{fig:inverse_str} shows the same comparison for the two
    stretched-exponential cases $c_t=1/4+(1/4)e^{-t^\beta}$ for $t\ge
    0$ with exponents $\beta=1/2$ and $\beta=1$, respectively. This
    time, the asymptotic equivalence
    $\rho_t\sim(1/\mu^3)\sum_{n>t}Q(n)$ is confirmed only for the
    subexponential case $\beta=1/2$. The exponential case $\beta=1$ is
    solved explicitly by the formulas $\rho_t=(1/4)e^{-t}$ and
    $(1/\mu^3)\sum_{n>t}Q(n)=(1/8)(\frac{1+e^{-1}}{2})^t$ for all
    $t\ge 0$, which shows that the decay rates of the autocovariance
    and of the associated waiting time distribution are different.

\section{Limit theorems}
\label{sec:empiricalmeans}

There are a number of limit theorems for the sequence $X:=\X$ that
follow from an underlying ergodic property. In this section we discuss
some of these limit theorems, proving at first ergodicity of a related
dynamical system. We refer to \cite{Walters} for basics of ergodic
theory. Recalling that $\mathscr{B}$ denotes the $\sigma$-field on
$\bin^\infty$ generated by the cylinder subsets, it is now convenient
to introduce the probability measure
$\prob_o[\,\cdot\,]:=\prob[X\in\cdot\,]$ on $\mathscr{B}$ and the
probability space $(\bin^\infty,\mathscr{B},\prob_o)$. We can deduce
almost sure convergence in $(\Omega,\mathscr{F},\prob)$ from almost
sure convergence in $(\bin^\infty,\mathscr{B},\prob_o)$. In fact,
since almost sure convergence is defined only in terms of probability
distributions \cite{Shiryaev}, if $G,G_1,G_2,\ldots$ are real
$\mathscr{B}$-measurable functions on $\bin^\infty$ such that
$\lim_{n\uparrow\infty}G_n(x)=G(x)$ for $\paex$, then
$\lim_{t\uparrow\infty}G_t(X)=G(X)$ $\prob$-a.s.. The same can be said
for convergence in mean. The reason to deal with the new probability
space $(\bin^\infty,\mathscr{B},\prob_o)$ is that it can be endowed
with a measure-preserving transformation. Such transformation is the
left-shift operator $\mathcal{T}$ that maps any binary sequence
$x:=\{x_t\}_{t\ge 1}\in\bin^\infty$ in
$\mathcal{T}x:=\{x_{t+1}\}_{t\ge 1}$.  The operator $\mathcal{T}$ is
measurable and, due to stationary of $X$, preserves measures, namely
$\prob_o[\mathcal{T}^{-1}\mathcal{B}]=\prob_o[\mathcal{B}]$ for each
$\mathcal{B}\in\mathscr{B}$.  Indeed, if $\mathcal{B}$ is an element
of $\mathscr{B}$, then
$\prob_o[\mathcal{T}^{-1}\mathcal{B}]=\prob[\mathcal{T}X\in\mathcal{B}]=\prob[\{X_{t+1}\}_{t\ge
    1}\in\mathcal{B}]=\prob[\X\in\mathcal{B}]=\prob_o[\mathcal{B}]$.
The transformation $\mathcal{T}$ is \textit{strong-mixing} in the
sense of ergodic theory, as stated by the following lemma that in
proved in Appendix \ref{proof:strongmixing}. The proof relies on Lemma
\ref{lem:covYZ}.
\begin{lemma}
\label{lem:strongmixing}
Assume that $p$ is aperiodic. Then, for all $\mathcal{A}$ and
$\mathcal{B}$ in $\mathscr{B}$
\begin{equation*}
  \lim_{t\uparrow\infty}\prob_o[\mathcal{A}\cap\mathcal{T}^{-t}\mathcal{B}]=\prob_o[\mathcal{A}]\,\prob_o[\mathcal{B}].
\end{equation*}
\end{lemma}

Due to strong-mixing, the transformation $\mathcal{T}$ is
\textit{ergodic} according to Corollary 1.14.2 of \cite{Walters},
which means that the only members $\mathcal{B}$ of $\mathscr{B}$ with
$\mathcal{T}^{-1}\mathcal{B}=\mathcal{B}$ satisfy
$\prob_o[\mathcal{B}]=0$ or $\prob_o[\mathcal{B}]=1$. We use
ergodicity to demonstrate an asymptotic equipartition property, which
justifies the principle of maximum entropy for model selection, and to
investigate the behavior of empirical means.

\subsection{Asymptotic equipartition property}

Description of data requires to select a statistical model, that is a
waiting time distribution $p$ once our framework is considered. One
tool for model selection is the \textit{maximum entropy principle}
\cite{Jaynes1,Jaynes2}, which would amount to pick the probability
distribution $p$ of finite mean that meets certain moment constraints
representing the available information and that maximizes the Shannon
entropy. The \textit{Shannon entropy} $\mathcal{H}(p)$ of $p$ is
defined by
\begin{equation*}
\mathcal{H}(p):=-\sum_{s\ge 1}p(s)\ln p(s)
\end{equation*}
with $0\ln 0:=0$. We point out that
$\mathcal{H}(p)\le\mu\ln\mu+(1-\mu)\ln(\mu-1)<\infty$ whenever
$\mu:=\sum_{s\ge 1}sp(s)<\infty$ since $-p(s)\ln p(s)\le
\mu^{-s}(\mu-1)^{s-1}-p(s)-p(s)\ln[\mu^{-s}(\mu-1)^{s-1}]$ for all
$s\ge 1$ by concavity. Actually, the largest value of the entropy is
$\mu\ln\mu+(1-\mu)\ln(\mu-1)$, which is attained if and only if
$p(s)=\mu^{-s}(\mu-1)^{s-1}$ for every $s$.  The Shannon entropy can
be derived axiomatically as a measure of uncertainty in the outcomes
of a random variable.  Instead, in this section we show that the
entropy $\mathcal{H}(p)$ of $p$ naturally arises as the answer to the
question ``how many typical sequences are there?''. In fact, we
demonstrate that there are about $e^{(t/\mu)\mathcal{H}(p)}$ typical
sequences of length $t$, each with probability about
$e^{-(t/\mu)\mathcal{H}(p)}$.  It follows that selecting the waiting
time distribution that maximizes the entropy means not excluding
possible sequences arbitrarily. If the only available information is
the mean $\mu$, then the maximum entropy prescription is the waiting
time distribution $p$ defined by $p(s)=\mu^{-s}(\mu-1)^{s-1}$ for all
$s$. According to Corollary \ref{Markovchain}, the binary sequence
$X$ associated with such $p$ is a sequence of i.i.d.\ binary random
variables with mean $1/\mu$.

Let us formalize and explain the above statements. Together with the
Shannon entropy of the waiting time distribution, we consider for each
$t\ge 1$ the Shannon entropy $\mathcal{H}(\pi_t)$ of the
finite-dimensional distribution $\pi_t$ of $X$:
\begin{equation*}
\mathcal{H}(\pi_t):=-\sum_{x_1\in\bin}\cdots\sum_{x_t\in\bin}\pi_t(x_1,\ldots,x_t)\ln\pi_t(x_1,\ldots,x_t).
\end{equation*}
Proposition \ref{prop:dist} and stationary and reversibility of $X$
give
\begin{align}
  \nonumber
  \mathcal{H}(\pi_t)&=\ln\mu-\Ex\Bigg[\prod_{n=1}^t(1-X_n)\Bigg]\ln\sum_{s\ge t}Q(s)-2\sum_{s=1}^t\Ex\Bigg[\prod_{n=1}^{s-1}(1-X_n)X_s\Bigg]\ln Q(s-1)\\
  \nonumber
  &-\sum_{s=1}^t(t-s)\,\Ex\Bigg[X_1\prod_{n=2}^s(1-X_n)X_{s+1}\Bigg]\ln p(s)\\
  \nonumber
  &=\ln\mu-\frac{1}{\mu}\sum_{s\ge t}Q(s)\ln\sum_{s\ge t}Q(s)-\frac{2}{\mu}\sum_{s=1}^t Q(s-1)\ln Q(s-1)\\
  \nonumber
  &-\frac{1}{\mu}\sum_{s=1}^t(t-s)\,p(s)\ln p(s),
\end{align}
where we have used the facts that
$\Ex[\prod_{n=1}^t(1-X_n)]=\prob[S_1>t]$,
$\Ex[\prod_{n=1}^{s-1}(1-X_n)X_s]=\prob[S_1=s]$, and
$\Ex[X_1\prod_{n=2}^s(1-X_n)X_{s+1}]=\prob[S_1=1,S_2=s]$ for all $s$.
After invoking the properties of Ces\'aro means and the dominated
converge theorem, this formula shows that the \textit{entropy rate} of
the sequence $X$ is
\begin{equation}
\lim_{t\uparrow\infty}\frac{\mathcal{H}(\pi_t)}{t}=\frac{\mathcal{H}(p)}{\mu}.
\label{entropyrate}
\end{equation}
Then, under ergodicity of the left-shift operator $\mathcal{T}$, the
Shannon-McMillan-Breiman theorem \cite{CoverThomas} yields for $\paex$
\begin{equation*}
\lim_{t\uparrow\infty}-\frac{1}{t}\ln\pi_t(x_1,\ldots,x_t)=\lim_{t\uparrow\infty}\frac{\mathcal{H}(\pi_t)}{t}=\frac{\mathcal{H}(p)}{\mu}.
\end{equation*}
These considerations, in combination with Lemma \ref{lem:strongmixing}
and the possibility to deduce almost sure convergence in
$(\Omega,\mathscr{F},\prob)$ from almost sure convergence in
$(\bin^\infty,\mathscr{B},\prob_o)$, prove the following asymptotic
equipartition property, which states that all long typical sequences
of length $t$ have roughly the same probability
$e^{-(t/\mu)\mathcal{H}(p)}$.
\begin{theorem}
\label{th:entropy}
Assume that $p$ is aperiodic. Then
\begin{equation*}
\lim_{t\uparrow\infty}-\frac{\mu}{t}\ln \pi_t(X_1,\ldots,X_t)=\mathcal{H}(p)~~~~~\prob\mbox{-a.s.}.
\end{equation*}
\end{theorem}

Theorem \ref{th:entropy} implies that there are about
$e^{(t/\mu)\mathcal{H}(p)}$ typical sequences of length $t$. This fact
is illustrated by the following corollary, whose proof is reported in
Appendix \ref{proof:typicalset}. A formal notion of typical set is
needed here.  Given a real number $\delta\in(0,1)$, according to
\cite{CoverThomas} we say that $\mathcal{X}\subseteq\bin^t$ is a
\textit{typical set} of length $t\ge 1$ if
$\prob[(X_1,\ldots,X_t)\in\mathcal{X}]\ge 1-\delta$.  We denote by
$|\mathcal{X}|$ the cardinality of a set $\mathcal{X}$.
\begin{corollary}
  \label{typicalset}
  Fix $\epsilon>0$ and assume that $p$ is aperiodic. The following
  conclusions hold for all sufficiently large $t$:
  \begin{enumerate}[(i)]
  \item there exists a typical set $\mathcal{X}_o$ of length $t$ such that
$|\mathcal{X}_o|\le e^{(t/\mu)[\mathcal{H}(p)+\epsilon]}$;
  \item any typical set $\mathcal{X}$ of length $t$ satisfies
$|\mathcal{X}|\ge e^{(t/\mu)[\mathcal{H}(p)-\epsilon]}$.
   \end{enumerate}
\end{corollary}

\subsection{Empirical means}

Applications to real data require to explain whether or not ensemble
averages can be estimated by means of time averages, also known as
empirical means.  If $G$ is a $\mathscr{B}$-measurable function on
$\bin^\infty$, then its \textit{empirical mean} up to time $t\ge 1$ is
the random variable
\begin{equation*}
\frac{1}{t}\sum_{n=1}^t G(X_n,X_{n+1},\ldots)=\frac{1}{t}\sum_{n=0}^{t-1} G(\mathcal{T}^nX).
\end{equation*}
Birkhoff ergodic theorem \cite{Walters} tells us that if the
left-shift operator $\mathcal{T}$ is ergodic and the expectation
$\int_{\bin^\infty} |G(x)|\,\prob_o[dx]=\Ex[|G(X)|]$ is finite, then
for $\paex$
\begin{equation*}
  \lim_{t\uparrow\infty}\frac{1}{t}\sum_{n=0}^{t-1} G(\mathcal{T}^nx)=\int_{\bin^\infty} G(x)\,\prob_o[dx]=\Ex[G(X)].
\end{equation*}
The convergence also holds in mean by Corollary 1.14.1 of
\cite{Walters}.  This way, Lemma \ref{lem:strongmixing} and the
possibility to export convergences from
$(\bin^\infty,\mathscr{B},\prob_o)$ to $(\Omega,\mathscr{F},\prob)$
result in the following law of large numbers, which gives a positive
answer to the possibility of estimating ensemble averages with time
averages. We stress that in most applications the observable $G$
depends only on a finite number of variables, so that $G$ is
automatically $\mathscr{B}$-measurable and bounded.
\begin{theorem}
\label{Birkhoff}
Let $G$ be a $\mathscr{B}$-measurable function on $\bin^\infty$ such
that $\Ex[|G(X)|]<\infty$. If $p$ is aperiodic, then
\begin{equation*}
\lim_{t\uparrow\infty}\frac{1}{t}\sum_{n=0}^{t-1} G(\mathcal{T}^nX)=\Ex[G(X)]~~~~~\prob\mbox{-a.s. and in mean}.
\end{equation*}
\end{theorem}

We deepen the study of empirical means investigating their normal
fluctuations.  Since $X$ is a regenerative process with independent
cycles \cite{Asmussen}, if $G$ depends on exactly one binary variable,
then the normal fluctuations of its empirical mean are described under
the optimal hypothesis $\Ex[S_2^2]=\sum_{s\ge 1}s^2p(s)<\infty$ by the
central limit theorem for cumulative processes with a regenerative
structure \cite{Asmussen}. It is worth noting that in such case the
empirical mean of $G$ up to time $t$ is a linear function of the
number of renewals by $t$, so that even its large fluctuations can be
completely characterized through well-established large deviation
principles for discrete-time renewal-reward processes
\cite{Marco1,Marco2}, which include the counting renewal process. To
deal with functions $G$ that involve more than one variable, and that
cannot be tackled by the standard limit theory of regenerative
processes with independent cycles, we resort to the theory of the
central limit theorem for stationary sequences \cite{Ibragimov}. To
begin with, we introduce the strong mixing coefficient $\alpha_t$ of
the sequence $X$, which measures the memory of the past on future
events $t$ time steps later. According to \cite{Bradley}, the
\textit{strong mixing coefficient} $\alpha_t$, or
\textit{$\alpha$-mixing coefficient}, of $X$ is defined for each $t\ge
1$ by
\begin{equation*}
  \alpha_t:=\sup_{m\ge 1}\,\adjustlimits\sup_{\mathcal{A}\in\mathscr{F}_1^m}\sup_{\mathcal{B}\in\mathscr{F}_{m+t}^\infty}
  \bigg\{\Big|\prob[\mathcal{A}\cap\mathcal{B}]-\prob[\mathcal{A}]\,\prob[\mathcal{B}]\Big|\bigg\}.
\end{equation*}
Here $\mathscr{F}_a^b$ is the $\sigma$-algebra generated by
$X_a,\ldots,X_b$ for $1\le a\le b\le\infty$.  The sequence $X$ is
\textit{strong mixing} in the sense of probability theory if
$\lim_{t\uparrow\infty}\alpha_t=0$.  The following proposition
provides an estimate of the $\alpha$-mixing coefficient of our
model. The proof is based on Lemma \ref{lem:covYZ} and is proposed in
Appendix \ref{proof:mixing}. Recall that $\rho_t:=\cov[X_1,X_{t+1}]$.
\begin{proposition}
  \label{prop:mixing}
  For each $t\ge 1$
\begin{equation*}
 \alpha_t\le 3\mu^2\sum_{n\ge t}|\rho_{n+1}-\rho_n|+4\mu^2\sum_{n\ge t}n\,|\rho_{n+1}-2\rho_n+\rho_{n-1}|.
\end{equation*}
\end{proposition}

Empirical means display normal fluctuations when the strong mixing
coefficient decays reasonably fast, and precisely when $\sum_{t\ge
  1}\alpha_t<\infty$. In fact, let us consider a bounded observable
$G$ and for each $n\ge 0$ let us set $Z_n:=G(\mathcal{T}^nX)$ for
brevity. Due to boundedness of $G$ Theorem \ref{Birkhoff} tells us
that $\lim_{t\uparrow\infty}(1/t)\sum_{n=0}^{t-1}Z_n=\Ex[Z_0]$
$\prob\mbox{-a.s.}$ if $p$ is aperiodic. The normal fluctuations of
$(1/t)\sum_{n=0}^{t-1}Z_n$ around $\Ex[Z_0]$ are described as follows
by Theorem 18.6.3 of \cite{Ibragimov}, which states the central limit
theorem for functionals of mixing sequences.
\begin{theorem}
\label{CLT}
Let $G$ be a bounded $\mathscr{B}$-measurable function on
$\bin^\infty$ and set $Z_n:=G(\mathcal{T}^nX)$ for $n\ge 0$.  Assume
that $\sum_{t\ge 1}\alpha_t<\infty$ and $\sum_{m\ge
  1}\Ex[|Z_0-\Ex[Z_0|\mathscr{F}_1^m]|]<\infty$. Then
\begin{equation*}
  v:=\cov[Z_0,Z_0]+2\sum_{n\ge 1}\cov[Z_0,Z_n]
\end{equation*}
is finite and non-negative, and provided that $v\ne 0$
  \begin{equation*}
    \lim_{t\uparrow\infty}\prob\Bigg[\frac{1}{\sqrt{vt}}\sum_{n=0}^{t-1} \big(Z_n-\Ex[Z_0]\big)\le z\Bigg]=\frac{1}{\sqrt{2\pi}}\int_{-\infty}^ze^{-\frac{1}{2}\zeta^2}d\zeta.
  \end{equation*}
\end{theorem}

We point out that the hypotheses of the theorem about $G$ are
automatically satisfied when $G$ depends only on a finite number of
variables, since in such case $\Ex[Z_0|\mathscr{F}_1^m]=Z_0$ for all
sufficiently large $m$.  The following lemma based on Proposition
\ref{prop:mixing} shows that the finiteness of the second moment of
the waiting time distribution suffices for $\sum_{t\ge
  1}\alpha_t<\infty$, thus implying the validity of the central limit
theorem. The proof is provided in Appendix \ref{proof:mixingcond}.
\begin{lemma}
  \label{lem:mixingcond}
  The two following statements are equivalent:
  \begin{enumerate}[(i)]
  \item $p$ is aperiodic and $\sum_{s\ge 1}s^2p(s)<\infty$;
  \item $\sum_{t\ge 1}t|\rho_t-\rho_{t-1}|<\infty$.
  \end{enumerate}
Either of them implies $\sum_{t\ge 1}\alpha_t<\infty$ and $\sum_{s\ge
  1}s^2p(s)=\mu+2\mu^3\sum_{t\ge 0}\rho_t$.
\end{lemma}

We stress that a coupling argument can show, without the need of an
explicit estimate of the $\alpha$-mixing coefficient, that an
aperiodic waiting time distribution satisfies $\sum_{s\ge 1}s^\gamma
p(s)<\infty$ for a real number $\gamma>1$ if and only if $\sum_{t\ge
  1}t^{\gamma-2}\alpha_t<\infty$ (see \cite{Mixing_bound}, Theorem
6.1).

\subsection{Empirical analyses}

Finally, we demonstrate the theory of empirical means through
estimation of the waiting time distribution and of the autocovariance
from data. The focus is on the possibility to identify their decay,
that is on the possibility to estimate $p(s)$ for large $s$ and
$\rho_\tau$ for large $\tau$. In order to avoid complications related
to nonlinear functions of empirical means, we imagine that $\mu$ is
known in advance. Importantly, we suppose that either $(i)$ or $(ii)$
of Lemma \ref{lem:mixingcond} hold, so that if $G$ is an observable
that depends on a finite number of variables, then both the law of
large numbers stated by Theorem \ref{Birkhoff} and the central limit
theorem of Theorem \ref{CLT} hold.

Fix an integer $s\ge 1$ and take
\begin{equation*}
  G(x):=\mu x_1\prod_{k=2}^s(1-x_k)x_{s+1}
\end{equation*}
for all $x\in\bin^\infty$.
\begin{figure}
  \centering
\includegraphics[height=5cm,width=10cm]{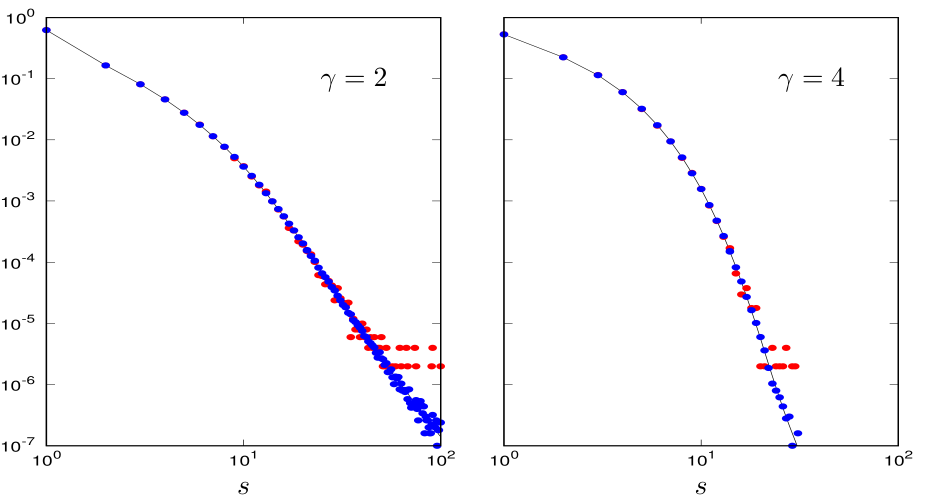}
\caption{Empirical estimates of $p(s)$ versus $s$ with data sequences
  of length $t=10^6$ (red) and $t=10^8$ (blue) generated by the models
  defined by $c_t=1/4+(1/4)(1+t)^{-\gamma}$ for $t\ge 0$ with
  $\gamma=2$ (left) and $\gamma=4$ (right). Black curves are the
  theoretical limits $p(s)$.}
\label{fig:Sim_wait_pol}
\vspace{0.8cm}
\includegraphics[height=5cm,width=10cm]{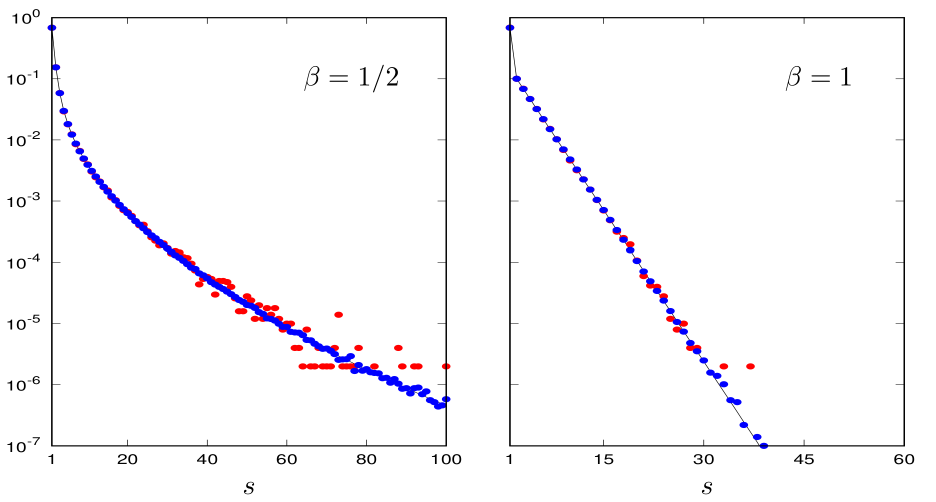}
\caption{Empirical estimates of $p(s)$ versus $s$ with data sequences
  of length $t=10^6$ (red) and $t=10^8$ (blue) generated by the models
  defined by $c_t=1/4+(1/4)e^{-t^\beta}$ for $t\ge 0$ with
  $\beta=1/2$ (left) and $\beta=1$ (right). Black curves are the
  theoretical limits $p(s)$.}
\label{fig:Sim_wait_str}
\end{figure}
Recalling that $Z_n:=G(\mathcal{T}^nX)$ for $n\ge 0$, we have
$\Ex[Z_0]=\mu\prob[S_1=1,S_2=s]=p(s)$, so that the empirical mean of
$G$ estimates the waiting time distribution at $s$. Simple
applications of Proposition \ref{prop:condind} show that
$\cov[Z_0,Z_n]=\mu p(s)-p^2(s)$ if $n=0$, $\cov[Z_0,Z_n]=-p^2(s)$ if
$1\le n<s$, and $\cov[Z_0,Z_n]=\mu^2p^2(s)\rho_{n-s}$ if $n\ge s$.
This way, the variance $v_s:=\cov[Z_0,Z_0]+2\sum_{n\ge
  1}\cov[Z_0,Z_n]$ introduced by Theorem \ref{CLT} turns out to be
\begin{align}
\nonumber
v_s&=\mu p(s)+(1-2s)p^2(s)+2\mu^2p^2(s)\sum_{n\ge 0}\rho_n\\
\nonumber
&=\mu p(s)-2sp^2(s)+\mu^{-1}p^2(s)\sum_{n\ge 1}n^2p(n).
\end{align}
We have made use of the identity $\sum_{s\ge
  1}s^2p(s)=\mu+2\mu^3\sum_{t\ge 0}\rho_t$ provided by Lemma
\ref{lem:mixingcond} to get at the second equality. If a data sequence
of length $t$ is given, then we expect to be able to estimate $p(s)$
for values of $s$ such that $\sqrt{v_s/t}\ll p(s)$. At large $s$, this
means values of $s$ such that $p(s)\gg \mu/t$ because $v_s\sim\mu
p(s)$.  Figure \ref{fig:Sim_wait_pol} shows results of estimation for
two data sequences of length $t=10^6$ and $t=10^8$, respectively, when
data are generated by the models of Figure \ref{fig:inverse_pol},
which correspond to $c_t=1/4+(1/4)(1+t)^{-\gamma}$ for $t\ge 0$ with
exponents $\gamma=2$ and $\gamma=4$. In such cases, the condition
$p(s)\gg \mu/t$ for proper estimation reads $s\ll 1.56\,t^{1/4}$ for
$\gamma=2$ and $s\ll 1.65\,t^{1/6}$ for $\gamma=4$.  Figure
\ref{fig:Sim_wait_str} reports the same estimation for the models of
Figure \ref{fig:inverse_str} defined by $c_t=1/4+(1/4)e^{-t^\beta}$
for $t\ge 0$ with exponents $\beta=1/2$ and $\beta=1$. The condition
$p(s)\gg \mu/t$ now becomes $s\ll\ln^2t$ for $\beta=1/2$ and $s\ll
2.63\,\ln t$ for $\beta=1$.

\begin{figure}
  \centering
\includegraphics[height=5cm,width=10cm]{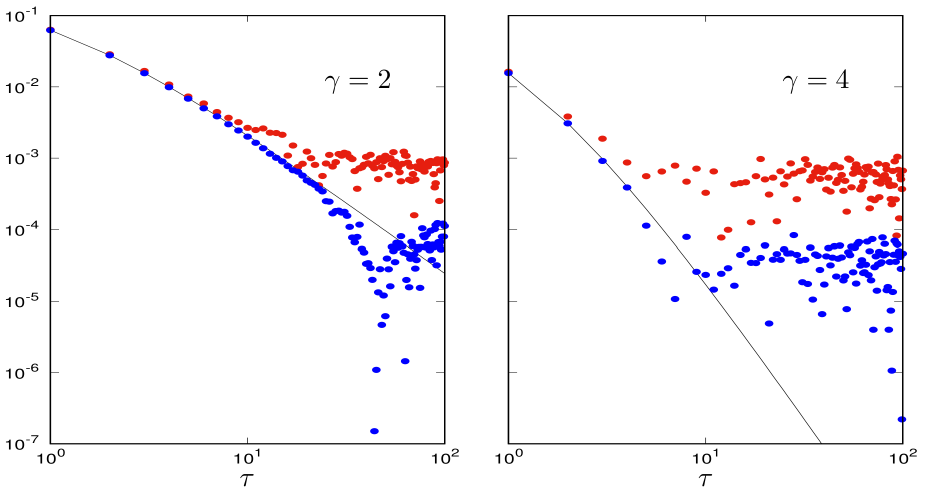}
\caption{Empirical estimates of $\rho_\tau$ versus $\tau$ with data
  sequences of length $t=10^6$ (red) and $t=10^8$ (blue) generated by
  the models defined by $c_t=1/4+(1/4)(1+t)^{-\gamma}$ for $t\ge 0$
  with $\gamma=2$ (left) and $\gamma=4$ (right). Black curves are the
  theoretical limits $\rho_\tau$.}
\label{fig:Sim_rho_pol}
\vspace{0.8cm}
\includegraphics[height=5cm,width=10cm]{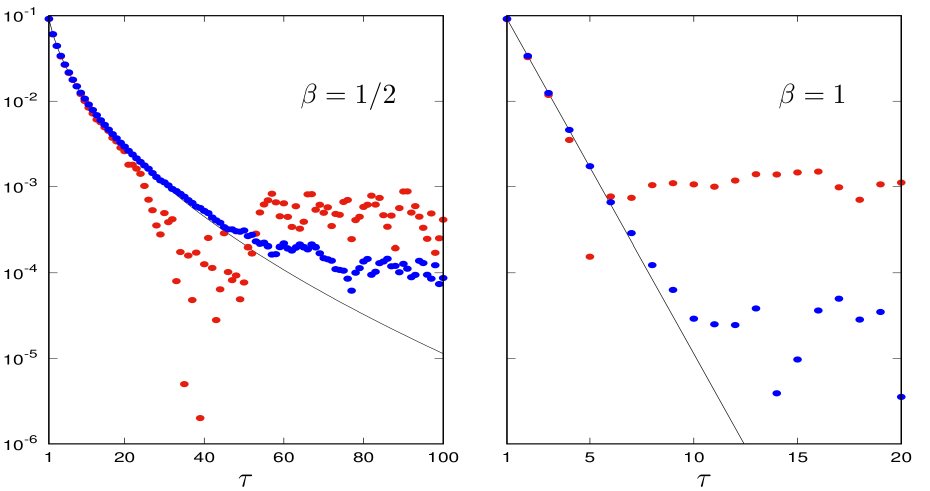}
\caption{Empirical estimates of $\rho_\tau$ versus $\tau$ with data
  sequences of length $t=10^6$ (red) and $t=10^8$ (blue) generated by
  the models defined by $c_t=1/4+(1/4)e^{-t^\beta}$ for $t\ge 0$ with
  $\beta=1/2$ (left) and $\beta=1$ (right). Black curves are the
  theoretical limits $\rho_\tau$.}
\label{fig:Sim_rho_str}
\end{figure}
Moving to the autocovariance, pick an integer $\tau\ge 0$ and consider
the function
\begin{equation}
  G(x):=x_1x_{\tau+1}-1/\mu^2
\label{G_rho}
\end{equation}
for all $x\in\bin^\infty$. As $\Ex[Z_0]=\rho_\tau$, the empirical mean
of $F$ estimates $\rho_\tau$. Once again, simple applications of
Proposition \ref{prop:condind} yield
$\cov[Z_0,Z_n]=\mu^2c_n^2c_{\tau-n}-c_\tau^2$ if $0\le n<\tau$ and
$\cov[Z_0,Z_n]=\mu^2 c_\tau^2\rho_{n-\tau}$ if $n\ge \tau$.  This way,
the variance $v_\tau:=\cov[Z_0,Z_0]+2\sum_{n\ge 1}\cov[Z_0,Z_n]$ reads
\begin{align}
  \nonumber
  v_\tau&=\begin{cases}
  \rho_0+2\sum_{n\ge 1}\rho_n & \mbox{if }\tau=0;\\
  c_\tau-c_\tau^2+2\sum_{n=1}^\tau(\mu^2c_n^2c_{\tau-n}-c_\tau^2)+2\mu^2c_\tau^2\sum_{n\ge 0}\rho_n & \mbox{if }\tau\ge 1
  \end{cases}\\
  &=\begin{cases}
  \mu^{-3}\sum_{n\ge 1}n^2p(n)-\mu^{-1} & \mbox{if }\tau=0;\\
  2\sum_{n=0}^\tau(\mu^2c_n^2c_{\tau-n}-c_\tau^2)+\mu^{-1}c_\tau^2\sum_{n\ge 1}n^2p(n)-c_\tau& \mbox{if }\tau\ge 1.
  \end{cases}
  \label{variance_mu}
\end{align}
Since
$\lim_{\tau\uparrow\infty}v_\tau=c_0^2(1-4c_0+3c_0^2)+8c_0^2\sum_{t\ge
  0}\rho_t+2\mu\sum_{t\ge 1}\rho_t^2=:\sigma^2$ with $\sigma>0$, we
expect that $\rho_\tau$ can be estimated with a data sequence of
length $t$ when $\tau$ satisfies $\rho_\tau\gg\sigma/\sqrt{t}$.
Figure \ref{fig:Sim_rho_pol} illustrates results of estimation for two
data sequences of length $t=10^6$ and $t=10^8$, respectively, when
data are generated by the models of Figures \ref{fig:inverse_pol} and
\ref{fig:Sim_wait_pol}. The condition $\rho_\tau\gg\sigma/\sqrt{t}$
explicitly is $\tau\ll 0.54\, t^{1/4}$ for $\gamma=2$ and $\tau\ll
0.77\, t^{1/8}$ for $\gamma=4$. Figure \ref{fig:Sim_rho_str} provides
the same results with reference to the models of Figures
\ref{fig:inverse_str} and \ref{fig:Sim_wait_str}. This time, the
condition $\rho_\tau\gg\sigma/\sqrt{t}$ is $\tau\ll 0.25\,\ln^2t$ for
$\beta=1/2$ and $\tau\ll 0.5\,\ln t$ for $\beta=1$. The comparison
between the waiting time distribution and the autocovariance shows
that the former is easier to estimate than the latter.

A final remark is in order. If $\mu$ is not known in advance, then its
inverse can be estimated through the empirical mean of the observable
$G(x):=x_1$ for all $x\in\bin^\infty$. Apart from an additive
constant, this observable is (\ref{G_rho}) with $\tau=0$. Thus, the
variance of its empirical mean exactly is $v_0$ given by
(\ref{variance_mu}).

\section{Conclusions}
\label{sec:conclusions}

We have explored the use of renewal processes to generate binary
sequences valued in $\{0,1\}$, where the symbol 1 marks a renewal
event. Focusing on stationary binary sequences corresponding to
delayed renewal processes, we have demonstrated the ability of the
model to account for subexponential autocovariances with special
attention to polynomial and stretched-exponential decays. Our model
performs at least as well as the algorithms proposed in \cite{Iter2}
and \cite{Beta1,Beta2} that seem to represent the state of the art,
but at variance with them, generating a binary sequence with our model
is a trivial task. Furthermore, our model is under full mathematical
control, and this fact allowed us to build a mathematical theory of
its asymptotic properties. In fact, in addition to shedding light on
the asymptotic behaviors of a large class of correlations and to
demonstrating an asymptotic equipartition property, we have developed
a theory for empirical means proving a law of large numbers and a
central limit theorem. The latter describes the typical fluctuations
of empirical means when the second moment of the waiting time
distribution is finite. We leave for future work the study of typical
fluctuations when the second moment of the waiting time distribution
is infinite. In such case, empirical means are expected to be not in
the Gaussian basin of attraction.  We also leave for future work the
investigation of their large fluctuations.

\section*{Acknowledgements and Funding}

The author is grateful to Alessandro Pelizzola for useful discussions.
This research was funded by Regione Puglia (Italy) through the
research programme REFIN - Research for Innovation (protocol code
A7B78549, project number UNIBA044).

\section*{Author declarations: Conflict of interest}
The author has no conflicts to disclose.

\section*{Author declarations: Data availability}
No datasets were used to support this study.

\appendix

\section{Proof of Theorem \ref{th:stationarity}}
\label{proof:stationarity}

A monotone class argument \cite{Shiryaev} shows that the sequence $X$
is stationary, namely that $\prob[\{X_{t+1}\}_{t\ge
    1}\in\mathcal{B}]=\prob[\X\in\mathcal{B}]$ for all
$\mathcal{B}\in\mathscr{B}$, if and only if $(X_2,\ldots,X_{t+1})$ is
distributed as $(X_1,\ldots,X_t)$ for every $t\ge 1$. If the process
$X$ is stationary, then for any $s\ge 2$
\begin{align}
  \nonumber
  \prob[T_1\ge s]=\prob\big[X_1=\cdots=X_{s-1}=0\big]&=\prob\big[X_2=\cdots=X_s=0\big]\\
  \nonumber
  &=\prob\big[X_1=0,X_2=\cdots=X_s=0\big]\\
  \nonumber
  &+\prob\big[X_1=1,X_2=\cdots=X_s=0\big]\\
  \nonumber
  &=\prob[T_1\ge s+1]+\prob[T_1=1,T_2\ge s+1],
\end{align}
which, due to the independence between $S_1=T_1$ and $S_2=T_2-T_1$, is
tantamount to
\begin{equation*}
\prob[S_1=s]=\prob[S_1=1]\,\prob[S_2\ge s].
\end{equation*}
This identity is trivial for $s=1$ and is valid for all $s\ge 1$ as a
consequence. In combination with the fact that $\sum_{s\ge
  1}\prob[S_1=s]=1$ and $\sum_{s\ge 1}\prob[S_2\ge s]=\Ex[S_2]$, it
implies $\prob[S_1=1]>0$ and $1=\prob[S_1=1]\,\Ex[S_2]$. Thus,
$\Ex[S_2]<\infty$ and $\Ex[S_2]\,\prob[S_1=s]=\prob[S_2\ge s]$ for
every $s\ge 1$.

Assume now that $\Ex[S_2]<\infty$ and
$\Ex[S_2]\,\prob[S_1=s]=\prob[S_2\ge s]$ for all $s\ge 1$. Let us
prove that $(X_2,\ldots,X_{t+1})$ is distributed as $(X_1,\ldots,X_t)$
for any given $t\ge 1$. Pick binary numbers $x_1,\ldots,x_t$. By
writing $X_{k+1}$ as $\mathds{1}_{\{k+1\in\{T_n\}_{n\ge 1}\}}$ and by
observing that $S_1$ is independent of $\{T_n-S_1\}_{n\ge 1}$ we find
\begin{align}
  \nonumber
  \prob\big[X_2=x_1,\ldots,X_{t+1}=x_t\big]&=\Ex\Bigg[\mathds{1}_{\{S_1=1\}}\prod_{k=1}^t\mathds{1}_{\big\{\mathds{1}_{\{k+1\in\{T_n\}_{n\ge
          1}\}}=x_k\big\}}\Bigg]\\
  \nonumber
  &+\Ex\Bigg[\mathds{1}_{\{S_1>1\}}\prod_{k=1}^t\mathds{1}_{\big\{\mathds{1}_{\{k+1\in\{T_n\}_{n\ge
          1}\}}=x_k\big\}}\Bigg]\\
  \nonumber
  &=\Ex\Bigg[\mathds{1}_{\{S_1=1\}}\prod_{k=1}^t\mathds{1}_{\big\{\mathds{1}_{\{k\in\{T_n-S_1\}_{n\ge 2}\}}=x_k\big\}}\Bigg]\\
  \nonumber
  &+\sum_{s\ge  1}\Ex\Bigg[\mathds{1}_{\{S_1=s+1\}}\prod_{k=1}^t\mathds{1}_{\big\{\mathds{1}_{\{k\in\{T_n-S_1+s\}_{n\ge
          1}\}}=x_k\big\}}\Bigg]\\
  \nonumber
  &=\prob[S_1=1]\,\Ex\Bigg[\prod_{k=1}^t\mathds{1}_{\big\{\mathds{1}_{\{k\in\{T_n-S_1\}_{n\ge 2}\}}=x_k\big\}}\Bigg]\\
  \nonumber
  &+\sum_{s\ge 1}\prob[S_1=s+1]\,\Ex\Bigg[\prod_{k=1}^t\mathds{1}_{\big\{\mathds{1}_{\{k\in\{T_n-S_1+s\}_{n\ge 1}\}}=x_k\big\}}\Bigg].
\end{align}
At this point, we use the formula
$\prob[S_1=s+1]=\prob[S_1=s]-\prob[S_1=1]\,\prob[S_2=s]$ valid for
each $s$, which is an immediate consequence of the hypothesis
$\Ex[S_2]\,\prob[S_1=s]=\prob[S_2\ge s]$. This way, by also noticing
that $\{T_n-S_1\}_{n\ge 1}$ is distributed as $\{T_n-S_1-S_2\}_{n\ge
  2}$, we get
\begin{align}
  \nonumber
  \prob\big[X_2=x_1,\ldots,X_{t+1}=x_t\big]&=\prob[S_1=1]\,\Ex\Bigg[\prod_{k=1}^t\mathds{1}_{\big\{\mathds{1}_{\{k\in\{T_n-S_1\}_{n\ge
          2}\}}=x_k\big\}}\Bigg]\\
  \nonumber
  &+\sum_{s\ge 1}\prob[S_1=s]\,\Ex\Bigg[\prod_{k=1}^t\mathds{1}_{\big\{\mathds{1}_{\{k\in\{T_n-S_1+s\}_{n\ge 1}\}}=x_k\big\}}\Bigg]\\
  \nonumber
  &-\prob[S_1=1]\sum_{s\ge 1}\prob[S_2=s]\,\Ex\Bigg[\prod_{k=1}^t\mathds{1}_{\big\{\mathds{1}_{\{k\in\{T_n-S_1-S_2+s\}_{n\ge 2}\}}=x_k\big\}}\Bigg].
\end{align}
The third term of the r.h.s.\ cancels with the first one since the
independence between $S_2$ and $\{T_n-S_1-S_2+s\}_{n\ge 2}$ shows that
the latter is an expanded version of the former. Then
\begin{align}
  \nonumber
  \prob\big[X_2=x_1,\ldots,X_{t+1}=x_t\big]&=
  \sum_{s\ge 1}\prob[S_1=s]\,\Ex\Bigg[\prod_{k=1}^t\mathds{1}_{\big\{\mathds{1}_{\{k\in\{T_n-S_1+s\}_{n\ge 1}\}}=x_k\big\}}\Bigg]\\
  \nonumber
  &=\Ex\Bigg[\prod_{k=1}^t\mathds{1}_{\big\{\mathds{1}_{\{k\in\{T_n\}_{n\ge 1}\}}=x_k\big\}}\Bigg]\\
  \nonumber
  &=\prob\big[X_1=x_1,\ldots,X_t=x_t\big].
\end{align}

\section{Proof of Proposition \ref{prop:dist}}
\label{proof:dist}

Pick an integer $t\ge 1$ and arbitrary numbers $x_1,x_2,\ldots,x_t$ in
$\bin$. If $x_1=x_2=\cdots=x_t=0$, then
\begin{align}
  \nonumber
  \pi_t(x_1,x_2,\ldots,x_t):&=\prob\big[X_1=x_1,X_2=x_2,\ldots,X_t=x_t\big]\\
  \nonumber
  &=\prob\big[T_1>t\big]=\prob\big[S_1>t\big]=\frac{1}{\mu}\sum_{s\ge t}Q(s).
\end{align}
If $x_1,x_2,\ldots,x_t$ account for exactly $N\ge 1$ ones in positions
$1\le i_1<i_2<\cdots<i_N\le t$, then
\begin{align}
  \nonumber
  \pi_t(x_1,x_2,\ldots,x_t)&=\prob\big[T_1=i_1,T_2=i_2,\ldots,T_N=i_N,T_{N+1}>t\big]\\
  \nonumber
  &=\prob\big[S_1=i_1,S_2=i_2-i_1,\ldots,S_N=i_N-i_{N-1},S_{N+1}>t-i_N\big]\\
  \nonumber
  &=\prob\big[S_1=i_1\big]\prod_{n=2}^N\prob\big[S_n=i_n-i_{n-1}\big]\,\prob\big[S_{N+1}>t-i_N\big]\\
  \nonumber
  &=\frac{1}{\mu}Q(i_1-1)\prod_{n=2}^Np(i_n-i_{n-1})\,Q(t-i_N).
\end{align}
These formulas together with the convention $0^0:=1$ show that
\begin{align}
  \nonumber
  \pi_t(x_1,x_2,\ldots,x_t)&=\frac{1}{\mu}\bigg[\sum_{s\ge t}Q(s)\bigg]^{\prod_{n=1}^t(1-x_n)}
  \prod_{i=1}^t\big[Q(i-1)\big]^{\prod_{n=1}^{i-1}(1-x_n)x_i}\\
  \nonumber
  &\,\,\cdot\,\,\prod_{i=1}^{t-1}\prod_{j=i+1}^t\big[p(j-i)\big]^{x_i\prod_{n=i+1}^{j-1}(1-x_n)x_j}\prod_{j=1}^t\big[Q(t-j)\big]^{x_j\prod_{n=j+1}^t(1-x_n)}.
\end{align}

Finally, by changing $i$ with $t-j+1$ and $j$ with $t-i+1$ in the
above products we realize that
\begin{equation*}
\pi_t(x_1,x_2,\ldots,x_t)=\pi_t(x_t,\ldots,x_2,x_1).
\end{equation*}

\section{Proof of Corollary \ref{Markovchain}}
\label{proof:Markovchain}

Thanks to the independence between $S_1$ and $S_2$, for all $t\ge 1$
we have
\begin{equation*}
\prob\big[X_1=1,X_2=0,\ldots,X_{t+1}=0\big]=\prob\big[T_1=1,T_2>t+1\big]=\prob\big[S_1=1,S_2>t\big]=\frac{Q(t)}{\mu}.
\end{equation*}
Thus, if $X$ is a sequence of i.i.d.\ binary random variables with
mean $1/\mu$, then for each $t$
\begin{equation*}
Q(t)=\bigg(1-\frac{1}{\mu}\bigg)^t.
\end{equation*}
Setting $\lambda:=1-1/\mu\in[0,1)$, this identity shows that
  $p(s+1)=\lambda^sp(1)$ for all $s\ge 1$. On the contrary, if
  $p(s+1)=\lambda^sp(1)$ for all $s\ge 1$ with some real number
  $\lambda\in[0,1)$, then Proposition \ref{prop:dist} demonstrates
    that $X$ is a sequence of i.i.d.\ random variables after some
    simple algebra.

Let us move to discuss Markovianity of order $M\ge 1$.  The sequence
$X$ is a Markov chain of order $M\ge 1$ if for all $t>M$ and
$x_1,\ldots,x_{t+1}$ such that $\prob[X_1=x_1,\ldots,X_t=x_t]>0$
\begin{equation*}
\prob\big[X_{t+1}=x_{t+1}\big|X_t=x_t,\ldots,X_1=x_1\big]=\prob\big[X_{t+1}=x_{t+1}\big|X_t=x_t,\ldots,X_{t-M+1}=x_{t-M+1}\big].
\end{equation*}
The constraint $\prob[X_1=x_1,\ldots,X_t=x_t]>0$ can be dropped by
multiplying both sides by $\prob[X_1=x_1,\ldots,X_t=x_t]$ and
$\prob[X_1=x_1,\ldots,X_M=x_M]$. In fact, the above condition is
tantamount to ask the for all $t>M$ and $x_1,\ldots,x_{t+1}$
\begin{align}
  \nonumber
  &\prob\big[X_1=x_1,\ldots,X_{t+1}=x_{t+1}\big]\,\prob\big[X_{t-M+1}=x_{t-M+1},\ldots,X_t=x_t\big]\\
\nonumber
  &=\prob\big[X_1=x_1,\ldots,X_t=x_t\big]\,\prob\big[X_{t-M+1}=x_{t-M+1},\ldots,X_{t+1}=x_{t+1}\big],
\end{align}
which, due to stationarity, reads   
\begin{align}
  \nonumber
  &\prob\big[X_1=x_1,\ldots,X_{t+1}=x_{t+1}\big]\,\prob\big[X_1=x_{t-M+1},\ldots,X_M=x_t\big]\\
  &=\prob\big[X_1=x_1,\ldots,X_t=x_t\big]\,\prob\big[X_1=x_{t-M+1},\ldots,X_{M+1}=x_{t+1}\big].
\label{Markov_proof_1}
\end{align}

By making the choice $x_1=1$ and $x_2=\cdots=x_{t+1}=0$ in
(\ref{Markov_proof_1}) we get for each $t>M$
\begin{equation*}
\prob\big[S_1=1,S_2>t\big]\,\prob\big[S_1>M\big]=\prob\big[S_1=1,S_2>t-1\big]\,\prob\big[S_1>M+1\big],
\end{equation*}
which, by appealing to the independence between $S_1$ and $S_2$ once
more, reduces to
\begin{equation}
  Q(t)\,Q(M)=Q(t-1)\,Q(M+1).
\label{Markov_necessary}
\end{equation}
This is a necessary condition for Markovianity of order $M\ge 1$.

In order to find a sufficient condition, we plug in
(\ref{Markov_proof_1}) the explicit expression of finite-dimensional
distributions given by Proposition \ref{prop:dist}. By equating the
factors that the two sides of the equation do not have in common we
reach the condition
\begin{equation}
  p(t)\sum_{s\ge M}Q(s)=Q(t-1)\,Q(M)
\label{Markov_sufficient}
\end{equation}
to be satisfied for every $t>M$.

Condition (\ref{Markov_necessary}) imposes that either $Q(M)=0$ or
$p(M+s+1)=\lambda^sp(M+1)$ for all $s\ge 1$ with
$\lambda:=Q(M+1)/Q(M)\in[0,1)$. Since $Q(M)=0$ implies $p(M+s+1)=0$
  for each $s\ge 0$, in any case we can state that
  (\ref{Markov_necessary}) entails that there exists a real number
  $\lambda\in[0,1)$ such that $p(M+s+1)=\lambda^sp(M+1)$ for all $s\ge
    1$. A waiting time distribution that fulfills such relation with
    some $\lambda\in[0,1)$ solves (\ref{Markov_sufficient}), as one
      can easily verify.

\section{Proof of Proposition \ref{prop:condind}}
\label{proof:condind}

Fix positive integers $s\le t$ and arbitrary binary numbers
$x_1,\ldots,x_t$. The proposition is trivial if $s=1$, $s=t$, or
$1<s<t$ and $x_s=0$. Then, assume $1<s<t$ and $x_s=1$. Since the
condition $X_s=1$ implies that $s\in\{T_m\}_{m\ge 1}$ and since
$T_1<T_2<\cdots$, we can write
\begin{align}
  \nonumber
  \prob\big[X_1=x_1,\ldots,X_t=x_t\big]&=\Ex\Bigg[\prod_{k=1}^{s-1}\mathds{1}_{\big\{\mathds{1}_{\{k\in\{T_n\}_{n\ge 1}\}}=x_k\big\}}\mathds{1}_{\{s\in\{T_m\}_{m\ge 1}\}}
    \prod_{k=s+1}^t\mathds{1}_{\big\{\mathds{1}_{\{k\in\{T_n\}_{n\ge 1}\}}=x_k\big\}}\Bigg]\\
  \nonumber
  &=\sum_{m\ge 1}\Ex\Bigg[\prod_{k=1}^{s-1}\mathds{1}_{\big\{\mathds{1}_{\{k\in\{T_n\}_{n=1}^m\}}=x_k\big\}}\mathds{1}_{\{T_m=s\}}
    \prod_{k=s+1}^t\mathds{1}_{\big\{\mathds{1}_{\{k-s\in\{T_n-T_m\}_{n\ge m}\}}=x_k\big\}}\Bigg].
\end{align}
For each $m\ge 1$, the sequence $\{T_n-T_m\}_{n\ge m}$ is independent
of $\{T_n\}_{n=1}^m$ and distributed as $\{T_n-T_1\}_{n\ge 1}\}$, so
that
\begin{align}
  \nonumber
  \prob\big[X_1=x_1,\ldots,X_t=x_t\big]
  &=\sum_{m\ge 1}\Ex\Bigg[\prod_{k=1}^{s-1}\mathds{1}_{\big\{\mathds{1}_{\{k\in\{T_n\}_{n=1}^m\}}=x_k\big\}}\mathds{1}_{\{T_m=s\}}
    \Bigg]\,\Ex\Bigg[\prod_{k=s}^t\mathds{1}_{\big\{\mathds{1}_{\{k-s\in\{T_n-T_1\}_{n\ge 1}\}}=x_k\big\}}\Bigg]\\
  \nonumber
  &=\Ex\Bigg[\prod_{k=1}^{s-1}\mathds{1}_{\big\{\mathds{1}_{\{k\in\{T_n\}_{n\ge 1}\}}=x_k\big\}}\mathds{1}_{\{s\in\{T_m\}_{m\ge 1}\}}
    \Bigg]\,\Ex\Bigg[\prod_{k=s}^t\mathds{1}_{\big\{\mathds{1}_{\{k-s\in\{T_n-T_1\}_{n\ge 1}\}}=x_k\big\}}\Bigg]\\
  \nonumber
  &=\prob\big[X_1=x_1,\ldots,X_s=x_s\big]\,\Ex\Bigg[\prod_{k=s}^t\mathds{1}_{\big\{\mathds{1}_{\{k-s\in\{T_n-T_1\}_{n\ge 1}\}}=x_k\big\}}\Bigg].
\end{align}
This identity proves the proposition since by taking the sum over
$x_1,\ldots,x_{s-1}$ we realize that
\begin{equation*}
\Ex\Bigg[\prod_{k=s}^t\mathds{1}_{\big\{\mathds{1}_{\{k-s\in\{T_n-T_1\}_{n\ge 1}\}}=x_k\big\}}\Bigg]=\prob\big[X_s=x_s,\ldots,X_t=x_t\big|X_s=1\big].
\end{equation*}

\section{Proof of Proposition \ref{prop:c}}
\label{proof:c}
Pick $t\ge 1$. The identity
$1=\sum_{s=1}^tX_{t-s+1}\prod_{k=t-s+2}^t(1-X_k)+\prod_{k=1}^t(1-X_k)$
results in the formula
$X_1=\sum_{s=1}^tX_1X_{t-s+1}\prod_{k=t-s+2}^t(1-X_k)$, which allows
us to write down
\begin{align}
\nonumber
c_t:=\Ex\big[X_1X_{t+1}\big]&=\sum_{s=1}^t\Ex\Bigg[X_1X_{t-s+1}\prod_{k=t-s+2}^t(1-X_k)X_{t+1}\Bigg]\\
\nonumber
&=\sum_{s=1}^t\Ex\Bigg[X_1\prod_{k=t-s+2}^t(1-X_k)X_{t+1}\Bigg|X_{t-s+1}=1\Bigg]\,\Ex[X_1].
\end{align}
This way, Proposition \ref{prop:condind} shows that
\begin{align}
\nonumber
c_t&=\sum_{s=1}^t\Ex\big[X_1\big|X_{t-s+1}=1\big]\,\Ex\Bigg[\prod_{k=t-s+2}^t(1-X_k)X_{t+1}\Bigg|X_{t-s+1}=1\Bigg]\,\Ex[X_1]\\
\nonumber
&=\sum_{s=1}^t\frac{\Ex[X_{t-s+1}\prod_{k=t-s+2}^t(1-X_k)X_{t+1}]}{\Ex[X_1]}\,c_{t-s}.
\end{align}
Finally, stationarity yields
\begin{align}
\nonumber
c_t&=\sum_{s=1}^t\frac{\Ex[X_1\prod_{k=2}^s(1-X_k)X_{s+1}]}{\Ex[X_1]}\,c_{t-s}\\
\nonumber
&=\sum_{s=1}^t\frac{\prob[T_1=1,T_2=s+1]}{\prob[T_1=1]}\,c_{t-s}=\sum_{s=1}^t\prob[S_2=s]\,c_{t-s}=\sum_{s=1}^tp(s)\,c_{t-s}.
\end{align}

\section{Proof of Lemma \ref{lem:covYZ}}
\label{proof:covYZ}

Set for brevity $Y:=f(X_1,\ldots,X_m)$,
$\Phi_i:=X_{m-i+1}\prod_{k=m-i+2}^m(1-X_k)$ for $1\le i\le m$, and
$\Psi_j(t):=\prod_{k=1}^{j-1}(1-X_{m+t+k-1})X_{m+t+j-1}$ for $j\ge 1$
and $t\ge 1$. We are going to show that for all $t\ge 1$
\begin{equation}
  \cov\big[Y,Z_{m+t}\big]=\sum_{i=1}^m\sum_{j=1}^n\cov\big[\Phi_i,\Psi_j(t)\big]\,
  \frac{\Ex[f(X_m,\ldots,X_1)\mathds{1}_{\{S_1=i\}}]}{\prob[S_1=i]}\,\frac{\Ex[\mathds{1}_{\{S_1=j\}}Z_1]}{\prob[S_1=j]}
  \label{covYZ_1}
\end{equation}
and, for every $i\ge 1$ and $j\ge 1$,
\begin{equation}
\cov\big[\Phi_i,\Psi_j(t)\big]=C_{i,j}(t).
 \label{covYZ_2}
\end{equation}

We verify (\ref{covYZ_1}) at first.  Pick $t\ge 1$. The identities
$1=\sum_{i=1}^m\Phi_i+\prod_{k=1}^m(1-X_k)$ and
$1=\sum_{j=1}^n\Psi_j(t)+\prod_{k=0}^{n-1}(1-X_{m+t+k})$ give
$Y=\sum_{i=1}^m Y\Phi_i$ and $Z_{m+t}=\sum_{j=1}^n \Psi_j(t)Z_{m+t}$
as $f(0,\ldots,0)=0$ and $g(0,\ldots,0)=0$ by hypothesis. Then, we can
write down the formula
\begin{equation}
\cov\big[Y,Z_{m+t}\big]=\sum_{i=1}^m\sum_{j=1}^n\bigg\{\Ex\big[Y\Phi_i\Psi_j(t)Z_{m+t}\big]
-\Ex\big[Y\Phi_i\big]\Ex\big[\Psi_j(t)Z_{m+t}\big]\bigg\}.
\label{covYZ_3}
\end{equation}
Let us manipulate $\Ex[Y\Phi_i\Psi_j(t)Z_{m+t}]$ for two given
indices $i\le m$ and $j\le n$. The condition $\Phi_i\ne 0$ implies
$X_{m-i+1}=1$ and $X_{m-i+2}=\cdots=X_m=0$.  This way, Proposition
\ref{prop:condind} yields
\begin{align}
\nonumber
\frac{\Ex[Y\Phi_i]}{\Ex[X_{m-i+1}]}&=\Ex\Big[f(X_1,\ldots,X_{m-i+1},0,\ldots,0)\Phi_i\Big|X_{m-i+1}=1\Big]\\
\nonumber
&=\Ex\Big[f(X_1,\ldots,X_{m-i+1},0,\ldots,0)\Big|X_{m-i+1}=1\Big]\,\Ex\Big[\Phi_i\Big|X_{m-i+1}=1\Big]\\
\nonumber
&=\frac{\Ex[f(X_1,\ldots,X_{m-i+1},0,\ldots,0)|X_{m-i+1}=1]\,\Ex[\Phi_i]}{\Ex[X_{m-i+1}]},
\end{align}
that is
\begin{equation*}
\Ex\big[Y\Phi_i\big]=\Ex\big[f(X_1,\ldots,X_{m-i+1},0,\ldots,0)\big|X_{m-i+1}=1\big]\,\Ex\big[\Phi_i\big].
\end{equation*}
Similarly, as the random quantity $\Psi_j(t)Z_{m+t}$ is a function of
$X_{m+t},\ldots,X_{m+t+n-1}$ only, we have
\begin{equation*}
\Ex\big[Y\Phi_i\Psi_j(t)Z_{m+t}\big]
=\Ex\big[f(X_1,\ldots,X_{m-i+1},0,\ldots,0)\big|X_{m-i+1}=1\big]\,\Ex\big[\Phi_i\Psi_j(t)Z_{m+t}\big].
\end{equation*}
By continuing with these arguments, since the condition $\Psi_j(t)\ne
0$ entails $X_{m+t}=\cdots=X_{m+t+j-2}=0$ and $X_{m+t+j-1}=1$, we
realize that
\begin{equation*}
\Ex\big[\Psi_j(t)Z_{m+t}\big]
=\Ex\big[\Psi_j(t)\big]\,\Ex\big[g(0,\ldots,0,X_{m+t+j-1},\ldots,X_{m+t+n-1})\big|X_{m+t+j-1}=1\big]
\end{equation*}
and
\begin{equation*}
\Ex\big[\Phi_i\Psi_j(t)Z_{m+t}\big]
=\Ex\big[\Phi_i\Psi_j(t)\big]\,\Ex\big[g(0,\ldots,0,X_{m+t+j-1},\ldots,X_{m+t+n-1})\big|X_{m+t+j-1}=1\big].
\end{equation*}
These four relations show that
\begin{equation}
  \Ex\big[Y\Phi_i\Psi_j(t)Z_{m+t}\big]=\frac{\Ex[Y\Phi_i]\,\Ex[\Phi_i\Psi_j(t)]\,\Ex[\Psi_j(t)Z_{m+t}]}
         {\Ex[\Phi_i]\,\Ex[\Psi_j(t)]}.
\label{covYZ_4}
\end{equation}
On the other hand, reversibility of $X$ gives
\begin{equation}
\Ex[Y\Phi_i]=\Ex\Bigg[f(X_m,\ldots,X_1)\prod_{k=1}^{i-1}(1-X_k)X_i\Bigg]=
\Ex\big[f(X_m,\ldots,X_1)\mathds{1}_{\{S_1=i\}}\big]
\label{covYZ_5}
\end{equation}
and
\begin{equation}
  \Ex[\Phi_i]=\prob[S_1=i].
\label{covYZ_6}
\end{equation}
Stationarity of $X$ implies
\begin{equation}
  \Ex\big[\Psi_j(t)Z_{m+t}\big]=\Ex\Bigg[\prod_{k=1}^{j-1}(1-X_k)X_jZ_1\Bigg]=\Ex\big[\mathds{1}_{\{S_1=j\}}Z_1\big]
\label{covYZ_7}
\end{equation}
and
\begin{equation}
  \Ex[\Psi_j(t)]=\prob[S_1=j].
\label{covYZ_8}
\end{equation}
Formula (\ref{covYZ_3}) in combination with (\ref{covYZ_4}),
(\ref{covYZ_5}), (\ref{covYZ_6}), (\ref{covYZ_7}), and (\ref{covYZ_8})
results in (\ref{covYZ_1}).

Let us move to (\ref{covYZ_2}), which we prove by induction with
respect to $t$. To begin with, let us verify that
$\cov[\Phi_i,\Psi_j(1)]=C_{i,j}(1)$ for each $i\ge 1$ and $j\ge 1$.
By recalling (\ref{covYZ_6}) and (\ref{covYZ_8}) and by invoking
stationarity of $X$ we find
\begin{align}
  \nonumber
  \cov\big[\Phi_i,\Psi_j(1)\big]&=\Ex\Bigg[X_{m-i+1}\prod_{k=m-i+2}^{m+j-1}(1-X_k)X_{m+j}\Bigg]-\Ex[\Phi_i]\,\Ex[\Psi_j(1)]\\
  \nonumber
  &=\Ex\Bigg[X_1\prod_{k=2}^{i+j-1}(1-X_k)X_{i+j}\Bigg]-\prob[S_1=i]\,\prob[S_1=j]\\
  \nonumber
  &=\prob[S_1=1,S_2=i+j-1]-\prob[S_1=i]\,\prob[S_1=j]\\
  &=\prob[S_1=1]\,p(i+j-1)-\prob[S_1=i]\,\prob[S_1=j].
  \label{covYZ_100}
\end{align}
On the other hand, by definition we have
\begin{align}
  \nonumber
  C_{i,j}(1)&=\sum_{u=1}^i\sum_{v=1}^j p(i-u)\,\rho_{u+v-1}\,p(j-v)\\
  \nonumber
  &=\sum_{u=1}^i\sum_{v=1}^j p(i-u)\,c_{u+v-1}\,p(j-v)-\sum_{u=1}^i\sum_{v=1}^j p(i-u)\,c_0^2\,p(j-v)\\
  \nonumber
  &=\sum_{u=0}^{i-1} p(u)\sum_{v=0}^{j-1}c_{i+j-u-1-v}\,p(v)-\frac{1}{\mu}\sum_{u=0}^{i-1}p(u)\,\frac{1}{\mu}\sum_{v=0}^{j-1}p(v).
\end{align}
Proposition \ref{prop:c} tells us that
$\sum_{v=1}^{i+j-u-1}c_{i+j-u-1-v}\,p(v)=c_{i+j-u-1}$ if $u\le
i-1$. Then, by recalling that $p(0):=-1$, we realize that
$\sum_{v=0}^{j-1}c_{i+j-u-1-v}\,p(v)=-\sum_{v=j}^{i+j-u-1}c_{i+j-u-1-v}\,p(v)$
for $u\le i-1$. The fact that $p(0):=-1$ also gives
$\sum_{u=0}^{i-1}p(u)=-\prob[S_2\ge i]$ and
$\sum_{v=0}^{j-1}p(v)=-\prob[S_2\ge j]$. Since $(1/\mu)\prob[S_2\ge
  i]=\prob[S_1=i]$ and $(1/\mu)\prob[S_2\ge j]=\prob[S_1=j]$ by
Theorem \ref{th:stationarity}, we obtain
\begin{align}
  \nonumber
  C_{i,j}(1)&=-\sum_{u=0}^{i-1}p(u)\sum_{v=j}^{i+j-u-1} c_{i+j-u-1-v}\,p(v)-\prob[S_1=i]\,\prob[S_1=j]\\
  \nonumber
  &=-\sum_{v=j}^{i+j-1}p(v)\sum_{u=0}^{i+j-v-1}c_{i+j-v-1-u}\,p(u)-\prob[S_1=i]\,\prob[S_1=j].
\end{align}
Another application of Proposition \ref{prop:c} yields
$\sum_{u=0}^{i+j-v-1}c_{i+j-v-1-u}\,p(u)=0$ if $v<i+j-1$ and
$\sum_{u=0}^{i+j-v-1}c_{i+j-v-1-u}\,p(u)=-c_0$ if $v=i+j-1$. Then
\begin{align}
  \nonumber
  C_{i,j}(1)&=p(i+j-1)c_0-\prob[S_1=i]\,\prob[S_1=j]\\
  &=\prob[S_1=1]\,p(i+j-1)-\prob[S_1=i]\,\prob[S_1=j].
  \label{covYZ_101}
\end{align}
Formulas (\ref{covYZ_100}) and (\ref{covYZ_101}) demonstrate that
$\cov[\Phi_i,\Psi_j(1)]=C_{i,j}(1)$ for each $i\ge 1$ and $j\ge 1$.

To conclude, let us show that if $\cov[\Phi_i,\Psi_j(t)]=C_{i,j}(t)$
for all positive indices $i$ and $j$ and some $t\ge 1$, then
$\cov[\Phi_i,\Psi_j(t+1)]=C_{i,j}(t+1)$. Fix $i\ge 1$ and $j\ge 1$. By
using the identity
$\Psi_j(t+1)=\Psi_{j+1}(t)+X_{m+t}\prod_{k=1}^{j-1}(1-X_{m+t+k})X_{m+t+j}$
and the induction hypothesis we get
\begin{align}
  \nonumber
  \cov\big[\Phi_i,\Psi_j(t+1)\big]&=C_{i,j+1}(t)+\cov\Bigg[\Phi_i,X_{m+t}\prod_{k=1}^{j-1}(1-X_{m+t+k})X_{m+t+j}\Bigg].
\end{align}
On the other hand, Proposition \ref{prop:condind} allows the factorization
\begin{equation*}
 \Ex\Bigg[\Phi_iX_{m+t}\prod_{k=1}^{j-1}(1-X_{m+t+k})X_{m+t+j}\Bigg]=\frac{\Ex[\Phi_iX_{m+t}]\,\Ex[X_{m+t}\prod_{k=1}^{j-1}(1-X_{m+t+k})X_{m+t+j}]}{\Ex[X_{m+t}]},
\end{equation*}
which results in
\begin{align}
  \nonumber
  \cov\Bigg[\Phi_i,X_{m+t}\prod_{k=1}^{j-1}(1-X_{m+t+k})X_{m+t+j}\Bigg]&=\frac{\Ex[\Phi_iX_{m+t}]\,\Ex[X_{m+t}\prod_{k=1}^{j-1}(1-X_{m+t+k})X_{m+t+j}]}{\Ex[X_{m+t}]}\\
  \nonumber
  &-\Ex[\Phi_i]\Bigg[X_{m+t}\prod_{k=1}^{j-1}(1-X_{m+t+k})X_{m+t+j}\Bigg]\\
  \nonumber
  &=\frac{\cov[\Phi_i,X_{m+t}]\,\Ex[X_{m+t}\prod_{k=1}^{j-1}(1-X_{m+t+k})X_{m+t+j}]}{\Ex[X_{m+t}]}.
\end{align}
Since $X_{m+t}=\Psi_1(t)$, the induction hypothesis and stationarity
of $X$ give
\begin{align}
  \nonumber
  \cov\Bigg[\Phi_i,X_{m+t}\prod_{k=1}^{j-1}(1-X_{m+t+k})X_{m+t+j}\Bigg]&=\frac{C_{i,1}(t)\,\Ex[X_1\prod_{k=2}^j(1-X_k)X_{j+1}]}{\Ex[X_1]}\\
  \nonumber
  &=\frac{C_{i,1}(t)\,\prob[S_1=1,S_2=j]}{\prob[S_1=1]}=C_{i,1}(t)\,p(j).
\end{align}
It follows that
\begin{align}
  \nonumber
  \cov\big[\Phi_i,\Psi_j(t+1)\big]&=C_{i,j+1}(t)+C_{i,1}(t)\,p(j)\\
  \nonumber
  &=\sum_{u=1}^i\sum_{v=1}^{j+1} p(i-u)\,\rho_{u+t+v-2}\,p(j-v+1)-\sum_{u=1}^ip(i-u)\,\rho_{u+t-1}\,p(j)\\
  \nonumber
  &=\sum_{u=1}^i\sum_{v=1}^j p(i-u)\,\rho_{u+t+1+v-2}\,p(j-v)=C_{i,j}(t+1).
\end{align}

\section{Proof of Corollary \ref{corol:stretched}}
\label{proof:stretched}

Set $\sigma(a):=\sup_{z\ge a}\{|z\ell'(z)/\ell(z)|\}$ and observe that
$\lim_{a\uparrow\infty}\sigma(a)=0$ by hypothesis.  The inequality
$|\ln\ell(b)-\ln\ell(a)|\le\int_a^b|\ell'(z)/\ell(z)|dz\le\sigma(a)(\ln
b-\ln a)$ gives for all sufficiently large $a\le b$ the bound
\begin{equation}
\bigg(\frac{a}{b}\bigg)^{\sigma(a)}\le\frac{\ell(b)}{\ell(a)}\le \bigg(\frac{b}{a}\bigg)^{\sigma(a)}.
\label{bound_str}
\end{equation}
This bound shows that $\ell$ is slowly varying since
$\lim_{a\uparrow\infty}\sigma(a)=0$. It also shows that $\beta
z^{\beta-1}\ell(z)/2\le(z+1)^\beta\ell(z+1)-z^\beta\ell(z)\le 2\beta
z^{\beta-1}\ell(z)$ for each $z$ large enough, so that
$\lim_{z\uparrow\infty}[(z+1)^\beta\ell(z+1)-z^\beta\ell(z)]=0$ due to
the fact that $\ell$ is dominated by polynomials according to
Proposition 1.3.6 of \cite{RV}. Then, $Q(t+1)\sim
e^{-(t+1)^\beta\ell(t+1)}\sim e^{-t^\beta\ell(t)}\sim Q(t)$.

Let $h$ be the continuously differentiable real function on a
neighborhood of infinity that maps $z$ in
$h(z):=z^{\beta-1}\ell(z)$. We now demonstrate that $-z^2h'(z)$ is
increasing to infinity with respect to $z$ and that $\int_\tau^\infty
e^{\frac{1}{2}z^2h'(z)}dz<\infty$ for some $\tau>0$. It follows that
$\{Q(t)\}_{t\ge 0}\in\mathscr{S}$ by the condition for
subexponentiality given in \cite{Chover}, so that Theorem
\ref{th:subexp} ensures us that $\rho_t\sim(1/\mu^3)\sum_{n>t}Q(n)$.
As $\lim_{z\uparrow\infty}z\ell'(z)/\ell(z)=0$ by hypothesis, the
formula
\begin{equation*}
  \frac{z^2h'(z)}{z^\beta\ell(z)}=\beta-1+\frac{z\ell'(z)}{\ell(z)}
\end{equation*}
yields
\begin{equation*}
  \lim_{z\uparrow\infty}\frac{z^2h'(z)}{z^\beta\ell(z)}=\beta-1<0.
\end{equation*}
Since $\ell$ is dominated by polynomials, this limit entails
$\lim_{z\uparrow\infty}-z^2h'(z)=\infty$ and at the same time shows
that $\int_\tau^\infty e^{\frac{1}{2}z^2h'(z)}dz<\infty$ for some
$\tau$ large enough. It remains to verify that $-z^2h'(z)$ is
increasing with respect to $z$.  The condition
$\lim_{z\uparrow\infty}z\ell'(z)/\ell(z)=0$ implies
$\lim_{z\uparrow\infty}z^2\ell''(z)/\ell(z)=0$ by L'H\^opital's rule
as $\lim_{z\uparrow\infty}z^2\ell''(z)/\ell(z)$ is assumed to exist.
Then, the formula
\begin{align}
  \nonumber
  \frac{[-z^2h'(z)]'}{z^{\beta-1}\ell(z)}&=\beta(1-\beta)-2\beta \frac{z\ell'(z)}{\ell(z)}-\frac{z^2\ell''(z)}{\ell(z)}\\
  \nonumber
\end{align}
results in
\begin{equation*}
  \lim_{z\uparrow\infty}\frac{[-z^2h'(z)]'}{z^{\beta-1}\ell(z)}=\beta(1-\beta)>0.
\end{equation*}
This suffices to state that $-z^2h'(z)$ is increasing with respect to
$z$.

To conclude, let us prove that
\begin{equation}
  \sum_{n>t}Q(n)\sim\frac{t^{1-\beta}}{\beta\ell(t)} e^{-t^\beta\ell(t)}.
\label{ultima_str}
\end{equation}
Pick $\epsilon\in(0,\beta\wedge 1-\beta)$ and recall that
$\lim_{a\uparrow\infty}\sigma(a)=0$ by hypothesis. The fact that
$Q(t)\sim e^{-t^\beta\ell(t)}$ and (\ref{bound_str}) imply that there
exists $t_o>0$ such that $(1-\epsilon)e^{-n^\beta\ell(n)}\le
Q(n)\le(1+\epsilon)e^{-n^\beta\ell(n)}$ and $t^\epsilon
n^{\beta-\epsilon}\ell(t)\le n^\beta\ell(n)\le
n^{\beta+\epsilon}(t+1)^{-\epsilon}\ell(t+1)$ for $n>t>t_o$. This way,
for every $t>t_o$ we get
\begin{align}
  \nonumber
  \sum_{n>t}Q(n)\le (1+\epsilon)\sum_{n>t}e^{-n^\beta\ell(n)}&\le  (1+\epsilon)\sum_{n>t}e^{-n^{\beta-\epsilon}t^\epsilon\ell(t)}\\
  \nonumber
  &\le(1+\epsilon)\int_{z>t}e^{-z^{\beta-\epsilon}t^\epsilon\ell(t)}dz\\
  \nonumber
  &=\frac{1+\epsilon}{\beta-\epsilon}\bigg[\frac{1}{t^\epsilon\ell(t)}\bigg]^{\frac{1}{\beta-\epsilon}}
  \int_{\zeta>t^\beta\ell(t)}\zeta^{\frac{1-\beta+\epsilon}{\beta-\epsilon}}e^{-\zeta}d\zeta
\end{align}
and
\begin{align}
  \nonumber
  \sum_{n>t}Q(n)\ge (1-\epsilon)\sum_{n>t}e^{-n^\beta\ell(n)}&\ge  (1-\epsilon)\sum_{n>t}e^{-n^{\beta+\epsilon}(t+1)^{-\epsilon}\ell(t+1)}\\
  \nonumber
  &\ge(1-\epsilon)\int_{z>t+1}e^{-z^{\beta+\epsilon}(t+1)^{-\epsilon}\ell(t+1)}dz\\
  \nonumber
  &=\frac{1-\epsilon}{\beta+\epsilon}\bigg[\frac{(t+1)^\epsilon}{\ell(t+1)}\bigg]^{\frac{1}{\beta+\epsilon}}
  \int_{\zeta>(t+1)^\beta\ell(t+1)}\zeta^{\frac{1-\beta-\epsilon}{\beta+\epsilon}}e^{-\zeta}d\zeta.
\end{align}
An integration by part shows that $a^\gamma
e^{-a}\le\int_{\zeta>a}\zeta^\gamma e^{-\zeta}d\zeta\le a^\gamma
e^{-a}/(1-\gamma/a)$ for every real numbers $\gamma>0$ and
$a>\gamma$. Then
\begin{equation*}
  \sum_{n>t}Q(n)\le
      \frac{1+\epsilon}{\beta-\epsilon}\frac{t^{1-\beta}}{\ell(t)}\frac{e^{-t^\beta\ell(t)}}{1-\frac{1-\beta+\epsilon}{\beta-\epsilon}\frac{1}{t^\beta\ell(t)}}
\end{equation*}
and
\begin{equation*}
  \sum_{n>t}Q(n)\ge \frac{1-\epsilon}{\beta+\epsilon}\frac{(t+1)^{1-\beta}}{\ell(t+1)}e^{-(t+1)^\beta\ell(t+1)}.
\end{equation*}
By recalling that $\ell$ is dominated by polynomials and that
$\lim_{z\uparrow\infty}[(z+1)^\beta\ell(z+1)-z^\beta\ell(z)]=0$ and by
observing that $\ell(t+1)\sim\ell(t)$ by the bound
$(1+1/t)^{-\sigma(t)}\le \ell(t+1)/\ell(t)\le (1+1/t)^{\sigma(t)}$
valid for sufficiently large $t$, we find
\begin{equation*}
  \frac{1-\epsilon}{\beta+\epsilon}\le\liminf_{t\uparrow\infty}\ell(t)t^{\beta-1}e^{t^\beta\ell(t)}\sum_{n>t}Q(n)\le
  \limsup_{t\uparrow\infty}\ell(t)t^{\beta-1}e^{t^\beta\ell(t)}\sum_{n>t}Q(n)\le\frac{1+\epsilon}{\beta-\epsilon}.
\end{equation*}
From here, we obtain (\ref{ultima_str}) by sending $\epsilon$ to zero.

\section{Proof of Theorem \ref{th:inverseproblem}}
\label{proof:inverseproblem}

The renewal equation defines a unique function $p$ according to the
scheme $p(1):=c_1/c_0$ and
$p(t+1):=c_{t+1}/c_0-\sum_{s=1}^tp(s)c_{t-s+1}/c_0$ for $t\ge 1$. We
must verify that $p(s)\ge 0$ for all $s\ge 1$ and that $\sum_{s\ge
  1}p(s)=1$.

We verify non-negativity of $p$ by induction. We have
$p(1):=c_1/c_0>0$. Pick $t\ge 1$ and assume that $p(s)\ge 0$ for every
$s\le t$. To say that $c$ is a Kaluza sequence means to say that
$\{c_n/c_{n-1}\}_{n\ge 1}$ is non-decreasing, so that
$c_{t-s+1}/c_{t-s}\le c_{t+1}/c_t$ for each $s\le t$. Then, the
renewal equation yields
\begin{equation*}
p(t+1)=c_{t+1}/c_0-\sum_{s=1}^t p(s)\,c_{t-s+1}/c_0\ge c_{t+1}[c_t-\sum_{s=1}^t p(s)\,c_{t-s}]/(c_0c_t)=0.
\end{equation*}

Let us now show that $p$ sums to 1. By combining non-negativity of $p$
with the renewal equation we get $\sum_{s=1}^np(s)\,c_{t-s}\le c_t$
for all $t\ge n\ge 1$. This bound gives $\sum_{s\ge 1}p(s)\le 1$ by
sending $t$ to infinity and by recalling that
$\lim_{t\uparrow\infty}c_t=c_0^2$ by hypothesis.  Then, the fact that
$\sum_{s\ge 1}p(s)<\infty$ allows us an application of the dominated
convergence theorem to the renewal equation, which shows that
$\sum_{s\ge 1}p(s)=1$ by using $\lim_{t\uparrow\infty}c_t=c_0^2$ once
again.

\section{Proof of Lemma \ref{lem:strongmixing}}
\label{proof:strongmixing}

Since $\mathscr{B}$ is the $\sigma$-algebra generated by cylinder
subsets of $\bin^\infty$, Theorem 1.17 of \cite{Walters} tells us that
it suffices to verify the lemma for cylinder subsets. Pick two
cylinder subsets $\mathcal{A}$ and $\mathcal{B}$ in $\bin^\infty$, so
that there exist integers $m\ge 1$ and $n\ge 1$ and sets
$A\subseteq\bin^m$ and $B\subseteq\bin^n$ such that
$\mathcal{A}=\{x\in\bin^\infty:(x_1,\ldots,x_m)\in A\}$ and
$\mathcal{B}=\{x\in\bin^\infty:(x_1,\ldots,x_n)\in B\}$. By
introducing the functions $f$ on $\bin^m$ and $g$ on $\bin^n$ defined
respectively by $f(x_1,\ldots,x_m):=\mathds{1}_{\{(x_1,\ldots,x_m)\in
  A\}}-\mathds{1}_{\{(0,\ldots,0)\in A\}}$ and
$g(x_1,\ldots,x_n):=\mathds{1}_{\{(x_1,\ldots,x_n)\in
  B\}}-\mathds{1}_{\{(0,\ldots,0)\in B\}}$, we can write for every
$t\ge 1$
\begin{align}
\nonumber
\prob_o\big[\mathcal{A}\cap\mathcal{T}^{-t-m+1}\mathcal{B}\big]-\prob_o\big[\mathcal{A}\big]\,\prob_o\big[\mathcal{B}\big]
&=\prob\big[X\in\mathcal{A},\mathcal{T}^{t+m-1}X\in\mathcal{B}]-\prob\big[X\in\mathcal{A}\big]\,\prob\big[X\in\mathcal{B}\big]\\
\nonumber
&=\prob\big[(X_1,\ldots,X_m)\in A,(X_{t+m},\ldots,X_{t+m+n-1})\in B]\\
\nonumber
&-\prob\big[(X_1,\ldots,X_m)\in A\big]\,\prob\big[(X_1,\ldots,X_n)\in B\big]\\
\nonumber
&=\cov\big[f(X_1,\ldots,X_m),g(X_{m+t},\ldots,X_{m+t+n-1})\big].
\end{align}
Since $f(0,\ldots,0)=g(0,\ldots,0)=0$, we can invoke Lemma
\ref{lem:covYZ} to conclude that for all $t\ge 1$
\begin{align}
  \nonumber
  &\!\!\prob_o\big[\mathcal{A}\cap\mathcal{T}^{-t-m+1}\mathcal{B}\big]-\prob_o\big[\mathcal{A}\big]\,\prob_o\big[\mathcal{B}\big]\\
&=\sum_{i=1}^m\sum_{j=1}^n C_{i,j}(t)\,
\frac{\Ex[\mathds{1}_{\{S_1=i\}}f(X_m,\ldots,X_1)]}{\prob[S_1=i]}\,\frac{\Ex[\mathds{1}_{\{S_1=j\}}g(X_1,\ldots,X_n)]}{\prob[S_1=j]},
\label{strongmixing_1}
\end{align}
where for $i\ge 1$ and $j\ge 1$ 
\begin{equation*}
  C_{i,j}(t):=\sum_{u=1}^i\sum_{v=1}^j p(i-u)\,\rho_{u+t+v-2}\,p(j-v)
\end{equation*}
with $p(0):=-1$. The only dependence on $t$ is contained in the
autocovariance. Formula \ref{strongmixing_1} proves the lemma since
aperiodicity of $p$ entails $\lim_{t\uparrow\infty}\rho_t=0$.

\section{Proof of Corollary \ref{typicalset}}
\label{proof:typicalset}

Set
$\mathcal{Y}_t:=\{(x_1,\ldots,x_t)\in\bin^t:|(\mu/t)\ln\pi_t(x_1,\ldots,x_t)+\mathcal{H}(p)|\le\epsilon\}$
for every integer $t\ge 1$.  Theorem \ref{th:entropy} yields
$\lim_{t\uparrow\infty}\prob[(X_1,\ldots,X_t)\in\mathcal{Y}_t]=1$, so
that $\prob[(X_1,\ldots,X_t)\in\mathcal{Y}_t]\ge 1-\delta$ for all
sufficiently large $t$. On the other hand, the bounds
$\sum_{(x_1,\ldots,x_t)\in\mathcal{Y}_t}\pi_t(x_1,\ldots,x_t)\le 1$
and $\pi_t(x_1,\ldots,x_t)\ge e^{-(t/\mu)[\mathcal{H}(p)+\epsilon]}$
for $(x_1,\ldots,x_t)\in\mathcal{Y}_t$ result in $|\mathcal{Y}_t|\le
e^{(t/\mu)[\mathcal{H}(p)+\epsilon]}$. Thus, $(i)$ is proved by taking
$\mathcal{X}_o=\mathcal{Y}_t$.

Let us verify $(ii)$.  Pick a typical set $\mathcal{X}$ of length $t$
and set
$\mathcal{Y}_t:=\{(x_1,\ldots,x_t)\in\bin^t:|(\mu/t)\ln\pi_t(x_1,\ldots,x_t)+\mathcal{H}(p)|\le\epsilon/2\}$
for $t\ge 1$. Since $\prob[(X_1,\ldots,X_t)\in\mathcal{X}]\ge
1-\delta$ by definition with $\delta\in(0,1)$ and
$\lim_{t\uparrow\infty}\prob[(X_1,\ldots,X_t)\in\mathcal{Y}_t]=1$ by
Theorem \ref{th:entropy}, for all sufficiently large $t$ we have
\begin{align}
\nonumber
 \prob\big[(X_1,\ldots,X_t)\in\mathcal{Y}_t\cap \mathcal{X}\big]&=\prob\big[(X_1,\ldots,X_t)\in\mathcal{Y}_t\big]+\prob\big[(X_1,\ldots,X_t)\in\mathcal{X}\big]\\
\nonumber
&-\prob\big[(X_1,\ldots,X_t)\in\mathcal{Y}_t\cup \mathcal{X}\big]\\
\nonumber
&\ge \prob\big[(X_1,\ldots,X_t)\in\mathcal{Y}_t\big]-\delta\ge e^{-(t/\mu)(\epsilon/2)}.
\end{align}
At the same time
\begin{align}
\nonumber
\prob\big[(X_1,\ldots,X_t)\in\mathcal{Y}_t\cap \mathcal{X}\big]&=\sum_{(x_1,\ldots,x_t)\in\mathcal{Y}_t\cap \mathcal{X}}\pi_t(x_1,\ldots,x_t)\\
\nonumber
&\le |\mathcal{X}|e^{-(t/\mu)[\mathcal{H}(p)-\epsilon/2]}.
\end{align}
These two bounds prove $(ii)$.

\section{Proof of Proposition \ref{prop:mixing}}
\label{proof:mixing}

Fix $t\ge 1$ and assume that $\sum_{n\ge t}|\rho_{n+1}-\rho_n|<\infty$
and $\sum_{n\ge t}n\,|\rho_{n+1}-2\rho_n+\rho_{n-1}|<\infty$,
otherwise there is nothing to prove. We verify first that for all
positive integers $m$ and $n$ and all sets
$\mathcal{A}\in\mathscr{F}_1^m$ and
$\mathcal{B}\in\mathscr{F}_{m+t}^{m+t+n-1}$
\begin{equation}
  \Big|\prob[\mathcal{A}\cap\mathcal{B}]-\prob[\mathcal{A}]\,\prob[\mathcal{B}]\Big|\le\sum_{i\ge 1}\sum_{j\ge 1}|C_{i,j}(t)|.
\label{mixing1}
\end{equation}
Coefficients $C_{i,j}(t)$ were introduced by Lemma \ref{lem:covYZ}.  A
monotone class argument \cite{Shiryaev} allows us to extend the bound
(\ref{mixing1}) to all $\mathcal{B}\in\mathscr{F}_{m+t}^\infty$, so
that $\alpha(t)\le\sum_{i\ge 1}\sum_{j\ge 1}|C_{i,j}(t)|$. Then we
demonstrate that
\begin{equation}
\sum_{i\ge 1}\sum_{j\ge 1}|C_{i,j}(t)|\le 3\mu^2\sum_{n\ge t}|\rho_{n+1}-\rho_n|+4\mu^2\sum_{n\ge t}n\,|\rho_{n+1}-2\rho_n+\rho_{n-1}|.
\label{mixing2}
\end{equation}

Let us verify (\ref{mixing1}). Pick $m\ge 1$, $n\ge 1$,
$\mathcal{A}\in\mathscr{F}_1^m$, and
$\mathcal{B}\in\mathscr{F}_{m+t}^{m+t+n-1}$. Since $\mathcal{A}$ is
measurable with respect to $\mathscr{F}_1^m$, there exists a set
$A\subseteq\bin^m$ such that
$\mathcal{A}=\{\omega\in\Omega:(X_1(\omega),\ldots,X_m(\omega))\in
A\}$. In the same way, there exists $B\subseteq\bin^n$ with the
property that
$\mathcal{B}=\{\omega\in\Omega:(X_{m+t}(\omega),\ldots,X_{m+t+n-1}(\omega))\in
B\}$. Set $f(x_1,\ldots,x_m):=\mathds{1}_{\{(x_1,\ldots,x_m)\in
  A\}}-\mathds{1}_{\{(0,\ldots,0)\in A\}}$ and
$g(x_1,\ldots,x_n):=\mathds{1}_{\{(x_1,\ldots,x_n)\in
  B\}}-\mathds{1}_{\{(0,\ldots,0)\in B\}}$. Then,
$f(0,\ldots,0)=g(0,\ldots,0)=0$ and Lemma \ref{lem:covYZ} yields
\begin{align}
  \nonumber
  \prob[\mathcal{A}\cap\mathcal{B}]-\prob[\mathcal{A}]\,\prob[\mathcal{B}]&=\prob\big[(X_1,\ldots,X_m)\in A,(X_{m+t},\ldots,X_{m+t+n-1})\in B\big]\\
  \nonumber
  &-\prob\big[(X_1,\ldots,X_m)\in A\big]\,\prob\big[(X_{m+t},\ldots,X_{m+t+n-1})\in B\big]\\ 
  \nonumber
  &=\cov\big[f(X_1,\ldots,X_m),g(X_{m+t},\ldots,X_{m+t+n-1})\big]\\
  \nonumber
  &=\sum_{i=1}^m\sum_{j=1}^n C_{i,j}(t)\,
\frac{\Ex[\mathds{1}_{\{S_1=i\}}f(X_m,\ldots,X_1)]}{\prob[S_1=i]}\,\frac{\Ex[\mathds{1}_{\{S_1=j\}}g(X_1,\ldots,X_n)]}{\prob[S_1=j]},
\end{align}
where for $i\ge 1$ and $j\ge 1$ 
\begin{equation*}
  C_{i,j}(t):=\sum_{u=1}^i\sum_{v=1}^j p(i-u)\,\rho_{u+t+v-2}\,p(j-v)
\end{equation*}
with $p(0):=-1$. As $f$ and $g$ can only take the values $-1$, $0$,
and $1$, this formula results in
\begin{equation*}
  \Big|\prob[\mathcal{A}\cap\mathcal{B}]-\prob[\mathcal{A}]\,\prob[\mathcal{B}]\Big|\le\sum_{i=1}^m\sum_{j=1}^n |C_{i,j}(t)|
  \le\sum_{i\ge 1}\sum_{j\ge 1}|C_{i,j}(t)|.
\end{equation*}

Let us move now to (\ref{mixing2}). Simple algebra shows that for all $i\ge 1$ and $j\ge 1$
\begin{align}
  \nonumber
  C_{i,j}(t)&=\rho_{i+t+j-2}-\sum_{u=1}^{i-1}p(i-u)\,\rho_{u+t+j-2}-\sum_{v=1}^{j-1}\rho_{i+t+v-2}\,p(j-v)\\
  \nonumber
  &+\sum_{u=1}^{i-1}\sum_{v=1}^{j-1} p(i-u)\,\rho_{u+t+v-2}\,p(j-v)\\
\nonumber
&=\rho_{i+t+j-2}\,Q(i-1)Q(j-1)\\
  \nonumber
  &+\sum_{u=1}^{i-1}p(u)\,\big(\rho_{i+t+j-2}-\rho_{i+t+j-2-u}\big)\,Q(j-1)+\sum_{v=1}^{j-1} Q(i-1)\,\big(\rho_{i+t+j-2}-\rho_{i+t+j-2-v}\big)\,p(v)\\
  \nonumber
  &+\sum_{u=1}^{i-1}\sum_{v=1}^{j-1} p(u)\,\big(\rho_{i+t+j-2}-\rho_{i+t+j-2-v}-\rho_{i+t+j-2-u}+\rho_{i+t+j-2-u-v}\big)\,p(v).
\end{align}
We recall that $Q$ is the tail of the probability distribution $p$.
This identity leads to the bound
\begin{equation}
  \sum_{i\ge 1}\sum_{j\ge 1}|C_{i,j}(t)|\le K_1+2K_2+K_3
\label{mixing7}
\end{equation}
with 
\begin{equation*}
K_1:=\sum_{i\ge 0}\sum_{j\ge 0}|\rho_{i+t+j}|\,Q(i)Q(j),
\end{equation*}
\begin{equation*}
K_2:=\sum_{i\ge 1}\sum_{j\ge 0}\sum_{u=1}^ip(u)\,\big|\rho_{i+t+j}-\rho_{i+t+j-u}\big|\,Q(j),
\end{equation*}
and
\begin{equation*}
K_3:=\sum_{i\ge 1}\sum_{j\ge 1}\sum_{u=1}^i\sum_{v=1}^j p(u)\,\big|\rho_{i+t+j}-\rho_{i+t+j-v}-\rho_{i+t+j-u}+\rho_{i+t+j-u-v}\big|\,p(v).
\end{equation*}
We address $K_1$, $K_2$, and $K_3$ separately.

To begin with, we stress that the condition $\sum_{n\ge
  t}|\rho_{n+1}-\rho_n|<\infty$ implies
$\lim_{n\uparrow\infty}\rho_n=0$. In fact, if $p$ is not aperiodic,
then there exists an integer $\tau>1$ such that $p(s)=0$ unless $s$ is
a multiple of $\tau$. The renewal equation gives $c_n=0$ unless $n$ is
a multiple of $\tau$, and the renewal theorem \cite{Erdos} shows that
$\lim_{k\uparrow\infty}c_{k\tau}=1/\mu$. In this situation,
$|\rho_{k\tau+1}-\rho_{k\tau}|=|c_{k\tau+1}-c_{k\tau}|=|c_{k\tau}|\ge
1/(2\mu)$ for all sufficiently large $k$, which contradicts the
hypothesis $\sum_{n\ge t}|\rho_{n+1}-\rho_n|<\infty$. Thus,
$\sum_{n\ge t}|\rho_{n+1}-\rho_n|<\infty$ requires that $p$ is
aperiodic, and hence that $\lim_{n\uparrow\infty}\rho_n=0$. The limit
$\lim_{n\uparrow\infty}\rho_n=0$ justifies the identity
$\rho_{i+t+j}=\sum_{n\ge i+t+j}(\rho_n-\rho_{n+1})$, which yields the
inequality $|\rho_{i+t+j}|\le|\sum_{n\ge t}|\rho_{n+1}-\rho_n|$, which
results in the bound
\begin{equation}
 K_1\le \mu^2\sum_{n\ge t}|\rho_{n+1}-\rho_n|.
\label{mixing8}
\end{equation}
As far as $K_2$ is concerned, by rearranging indices we get
\begin{equation*}
K_2=\sum_{u\ge 1}\sum_{l\ge t}p(u)\,|\rho_{l+u}-\rho_l|\sum_{j=0}^{l-t}Q(j)\le\mu\sum_{u\ge 1}\sum_{l\ge t}p(u)\,|\rho_{l+u}-\rho_l|.
\end{equation*}
At this point, the bound
$|\rho_{l+u}-\rho_l|\le\sum_{n=l}^{l+u-1}|\rho_{n+1}-\rho_n|$ shows
that
\begin{align}
  \nonumber
  K_2&\le\mu\sum_{u\ge 1}\sum_{l\ge t}p(u)\sum_{n=l}^{l+u-1}|\rho_{n+1}-\rho_n|\\
  \nonumber
  &=\mu\sum_{n\ge t}\sum_{u\ge 1}(n-t+1\wedge u)\,p(u)\,|\rho_{n+1}-\rho_n|\\
  &\le\mu\sum_{n\ge t}\sum_{u\ge 1}up(u)\,|\rho_{n+1}-\rho_n|=\mu^2\sum_{n\ge t}|\rho_{n+1}-\rho_n|.
\label{mixing9}
\end{align}
Finally, let us consider $K_3$. By rearranging indices we can write
\begin{align}
  \nonumber
 K_3&= \sum_{l\ge t}\sum_{u\ge 1}\sum_{v\ge 1}(l-t+1)\,p(u)\,\big|\rho_{l+u+v}-\rho_{l+u}-\rho_{l+v}+\rho_l\big|\,p(v)\\
\nonumber
&\le 2\sum_{l\ge t}l\sum_{u\ge 1}\sum_{v\ge u}p(u)\,\big|\rho_{l+u+v}-\rho_{l+u}-\rho_{l+v}+\rho_l\big|\,p(v).
\end{align}
In order to go one step further, setting
$\Delta_n:=\rho_{n+1}-2\rho_n+\rho_{n-1}$ for brevity, we resort to
the identity
\begin{equation*}
  \rho_{l+u+v}-\rho_{l+u}-\rho_{l+v}+\rho_l=\sum_{k=v+1}^{u+v}(u+v-k)\,\Delta_{k+l}+\sum_{k=1}^v u\wedge k\,\Delta_{k+l},
\end{equation*}
which holds for $u\le v$ as one can easily verify. This identity gives
\begin{equation}
  K_3\le 2M_1+2M_2
 \label{mixing10}
\end{equation}
with
\begin{equation*}
  M_1:=\sum_{l\ge t}l\sum_{u\ge 1}\sum_{v\ge u}p(u)\sum_{k=v+1}^{u+v}(u+v-k)\,|\Delta_{k+l}|\,p(v)
\end{equation*}
and
\begin{equation*}
  M_2:=\sum_{l\ge t}l\sum_{u\ge 1}\sum_{v\ge u}p(u)\sum_{k=1}^v u\wedge k\,|\Delta_{k+l}|\,p(v).
\end{equation*}
By rearranging indices we can write
\begin{align}
  \nonumber
  M_1&=\sum_{n\ge t+2}|\Delta_n|\sum_{u=1}^{n-t-1}\sum_{v=u}^{n-t-1}p(u)\,p(v)\sum_{l=t\vee (n-u-v)}^{n-v-1}l(u+v-n+l)\\
  \nonumber
  &\le\sum_{n\ge t+2}|\Delta_n|\sum_{u=1}^{n-t-1}\sum_{v=u}^{n-t-1}p(u)\,p(v)\,nu\big[n-v-t\vee (n-u-v)\big].
\end{align}
Under the condition $u\le v$, if $n-u-v\ge t$, then $n-v-t\vee (n-u-v)=u\le v$ and if $n-u-v<t$, then
$n-v-t\vee (n-u-v)=n-v-t<u\le v$. We thus have
\begin{equation}
  M_1\le\sum_{n\ge t+2}|\Delta_n|\sum_{u=1}^{n-t-1}\sum_{v=u}^{n-t-1}p(u)\,p(v)\,nuv\le\mu^2\sum_{n\ge t}n\,|\Delta_n|.
\label{mixing11}
\end{equation}
Moving to $M_2$, by rearranging indices we find
\begin{align}
  \nonumber
  M_2&\le\sum_{l\ge t}l\sum_{u\ge 1}\sum_{v\ge u}up(u)\sum_{k=1}^v|\Delta_{k+l}|\,p(v)\\
  \nonumber
  &=\sum_{n\ge t+1}|\Delta_n|\sum_{u\ge 1}\sum_{v\ge u}up(u)\,p(v)\sum_{l=t\vee (n-v)}^{n-1}l\\
  \nonumber
  &\le\sum_{n\ge t+1}|\Delta_n|\sum_{u\ge 1}\sum_{v\ge u}up(u)\,p(v)\,n\big[n-t\vee (n-v)\big].
\end{align}
If $n-v\ge t$, then $n-t\vee (n-v)=v$ and if $n-v<t$, then
$n-t\vee(n-v)=n-t<v$. It follows that
\begin{equation}
  M_2\le\sum_{n\ge t+1}|\Delta_n|\sum_{u\ge 1}\sum_{v\ge u}up(u)\,p(v)\,nv\le\mu^2\sum_{n\ge t}n\,|\Delta_n|.
\label{mixing12}
\end{equation}
Bound (\ref{mixing2}) follows by combining (\ref{mixing7}) with
(\ref{mixing8}), (\ref{mixing9}), (\ref{mixing10}), (\ref{mixing11}),
and (\ref{mixing12}).

\section{Proof of Lemma \ref{lem:mixingcond}}
\label{proof:mixingcond}

Assume that $p$ is aperiodic. To begin with, we demonstrate that there
exists a real sequence $\{\gamma_t\}_{t\ge 0}$ such that $\sum_{t\ge
  0}|\gamma_t|<\infty$ and
\begin{equation}
\frac{1}{\sum_{t\ge 0}Q(t)z^t}=\sum_{t\ge 0}\gamma_t z^t
\label{expansion}
\end{equation}
for all complex numbers $z$ in the open unit disk, $Q$ being the tail
of the waiting time distribution $p$. Consider the function $f$
defined by $f(z):=\sum_{t\ge 0}Q(t)z^t$ for $|z|\le 1$. We have
$1-\sum_{s\ge 1}p(s)z^s=(1-z)f(z)$ for all $|z|\le 1$. We claim that
$f$ has no zeros for $|z|\le 1$. In fact, $|z|<1$ entails $|\sum_{s\ge
  1}p(s)z^s|<1$, so that any zeros of $f$ must occur on the circle
$|z|=1$. The point $z=1$ cannot be a zero since $f(1)=\mu\ne 0$. If
$f(e^{{\rm i}\theta})=0$ for some $\theta\in(0,2\pi)$, then
$\sum_{s\ge 1}p(s)e^{{\rm i}s\theta}=1$, which gives $\cos(s\theta)=1$
for all $s\ge 1$ such that $p(s)>0$. But this is impossible because
$p$ is aperiodic. Then, the function that maps $z$ in $1/f(z)$ has no
singularities in the open unit disk and is continuous in the closed
unit disk. As a consequence, for $|z|<1$ it can be expanded in a power
series as $1/f(z)=\sum_{t\ge 0}\gamma_t z^t$ with $\gamma_t$ defined
for every $t\ge 0$ by
\begin{equation*}
\gamma_t:=\frac{1}{2\pi{\rm i}}\int_{|z|=r}\frac{z^{-t-1}}{f(z)}dz=\frac{1}{2\pi r^t}\int_0^{2\pi}\frac{e^{-{\rm i}t\theta}}{f(re^{{\rm i}\theta})}d\theta.
\end{equation*}
Here $r$ is any number in $(0,1)$. As $1/f$ is bounded on the closed
unit disk by continuity, we can invoke the dominated convergence
theorem to set $r=1$ in the last integral:
\begin{equation}
\gamma_t=\frac{1}{2\pi}\int_0^{2\pi}\frac{e^{-{\rm i}t\theta}}{f(e^{{\rm i}\theta})}d\theta.
\label{Wiener}
\end{equation}
At the same time, since $f(e^{{\rm i}\theta})=\sum_{t\ge 0}Q(t)e^{{\rm
    i}t\theta}$ converges absolutely and has no zeros for real values
of $\theta$, Theorem 18.21 of \cite{Rudin} states that $1/f(e^{{\rm
    i}\theta})=\sum_{t=-\infty}^\infty\xi_te^{{\rm i}t\theta}$ for all
$\theta\in[0,2\pi]$ with coefficients that fulfills
$\sum_{t=-\infty}^\infty|\xi_t|<\infty$. By plugging this expansion in
(\ref{Wiener}) we get $\gamma_t=\xi_t$ for every $t\ge 0$, and hence
$\sum_{t\ge 0}|\gamma_t|<\infty$.

We now observe that by taking a Cauchy product, formula
(\ref{expansion}) shows that the sequence $\{\gamma_t\}_{t\ge 0}$
solves the problem $\gamma_0=1$ and
$\gamma_t=-\sum_{s=1}^tQ(s)\gamma_{t-s}$ for all $t\ge 1$. Simple
manipulations of the renewal equation demonstrate that the same
problem is solved by the sequence $\{\gamma_t'\}_{t\ge 0}$ with
entries $\gamma_0'=1$ and $\gamma_t'=\mu(\rho_t-\rho_{t-1})$ for $t\ge
1$. Since this problem has a unique solution, we get
$\gamma_t=\mu(\rho_t-\rho_{t-1})$ for $t\ge 1$ and
\begin{equation*}
\frac{1}{\sum_{t\ge 0}Q(t)z^t}=1+\mu\sum_{t\ge 1}(\rho_t-\rho_{t-1}) z^t
\end{equation*}
for every $|z|<1$.  Derivative with respect to $z$ gives
\begin{equation}
\frac{\sum_{t\ge 1} tQ(t)z^t}{[\sum_{t\ge 0}Q(t)z^t]^2}=\mu\sum_{t\ge 1}t(\rho_{t-1}-\rho_t) z^t.
\label{expansion1}
\end{equation}
This formula is what we need to prove the equivalence between points
$(i)$ and $(ii)$ of the lemma. In fact, if $p$ is aperiodic and
$\sum_{s\ge 1}s^2p(s)<\infty$, then $1/\sum_{t\ge 0}Q(t)z^t=\sum_{t\ge
  0}\gamma_tz^t$ with $\sum_{t\ge 0}|\gamma_t|<\infty$ and $\sum_{t\ge
  1} tQ(t)=\sum_{s\ge 1}(1/2)(s-1)sp(s)<\infty$. This way, since the
Cauchy product of absolutely convergent series is absolutely
convergent, by expanding the l.h.s.\ of (\ref{expansion1}) in a power
series and by making a comparison with the r.h.s.\ we realize that
$\sum_{t\ge 1}t|\rho_t-\rho_{t-1}|<\infty$. If instead $\sum_{t\ge
  1}t|\rho_t-\rho_{t-1}|<\infty$, then $p$ is aperiodic as we have
seen in the proof of Proposition \ref{prop:mixing}, so that
(\ref{expansion1}) holds. It follows that $\sum_{t=1}^n
tQ(t)\le\mu^3\sum_{t\ge 1}t|\rho_t-\rho_{t-1}|<\infty$ for all
positive $n$, which suffices to prove that $\sum_{s\ge
  1}s^2p(s)<\infty$.

Suppose now that $p$ is aperiodic and that $\sum_{s\ge
  1}s^2p(s)<\infty$. Let us demonstrate that $\sum_{t\ge
  1}\alpha_t<\infty$. Proposition \ref{prop:mixing} tells us that
$\sum_{t\ge 1}\alpha_t\le3\mu^2\sum_{t\ge
  1}t|\rho_t-\rho_{t-1}|+4\mu^2\sum_{t\ge
  1}t^2|\rho_{t+1}-2\rho_t+\rho_{t-1}|$. We already know that
$\sum_{t\ge 1}t|\rho_t-\rho_{t-1}|<\infty$ and it remains to verify
that $\sum_{t\ge 1}t^2|\rho_{t+1}-2\rho_t+\rho_{t-1}|<\infty$. By
taking the derivative of (\ref{expansion1}) with respect to $z$ we get
\begin{equation}
  \frac{\sum_{t\ge 1} t^2 Q(t)z^t}{[\sum_{t\ge 0}Q(t)z^t]^2}-2\frac{[\sum_{t\ge 1} tQ(t)z^t]^2}{[\sum_{t\ge 0}Q(t)z^t]^3}
  =\mu\sum_{t\ge 1}t^2(\rho_{t-1}-\rho_t) z^t.
\label{expansion2}
\end{equation}
A shift of the index of the last series yields
\begin{equation}
  \frac{\sum_{t\ge 1} t^2 Q(t)z^{t-1}}{[\sum_{t\ge 0}Q(t)z^t]^2}-2z\frac{[\sum_{t\ge 1} tQ(t)z^{t-1}]^2}{[\sum_{t\ge 0}Q(t)z^t]^3}
  =\mu\sum_{t\ge 0}(t+1)^2(\rho_t-\rho_{t+1}) z^t.
  \label{expansion3}
\end{equation}
By subtracting (\ref{expansion3}) from (\ref{expansion2}) we reach the
result
\begin{align}
  \nonumber
  &\frac{\sum_{t\ge 1} t^2 p(t+1)z^t+\sum_{t\ge 1} (1-2t)Q(t)z^{t-1}}{[\sum_{t\ge 0}Q(t)z^t]^2}
  +2z(1-z)\frac{[\sum_{t\ge 1} tQ(t)z^{t-1}]^2}{[\sum_{t\ge 0}Q(t)z^t]^3}\\
  &=\mu\sum_{t\ge 1}t^2(\rho_{t-1}-2\rho_t+\rho_{t+1}) z^t+\mu\sum_{t\ge 1}(1-2t)(\rho_{t-1}-\rho_t) z^{t-1}.
 \label{expansion4}
\end{align}
As before, using $1/\sum_{t\ge 0}Q(t)z^t=\sum_{t\ge 0}\gamma_tz^t$ and
bearing in mind that the Cauchy product of absolutely convergent
series is absolutely convergent, by expanding the l.h.s.\ of
(\ref{expansion4}) in a power series and by making a comparison with
the r.h.s.\ we find $\sum_{t\ge
  1}t^2|\rho_{t-1}-2\rho_t+\rho_{t+1}|<\infty$.

At last, let us verify the formula $\sum_{s\ge
  1}s^2p(s)=\mu+2\mu^3\sum_{t\ge 0}\rho_t$. If $p$ is aperiodic and
$\sum_{s\ge 1}s^2p(s)<\infty$, or equivalently if $\sum_{t\ge
  1}t|\rho_t-\rho_{t-1}|<\infty$, then we can apply Abel's theorem to
(\ref{expansion1}) to obtain $\sum_{s\ge
  1}s^2p(s)=\mu+2\mu^3\sum_{t\ge 1}t(\rho_{t-1}-\rho_t)$. We have
$\sum_{t\ge 0}|\rho_t|<\infty$ since
$\rho_t=\sum_{n>t}(\rho_{n-1}-\rho_n)$ as
$\lim_{n\uparrow\infty}\rho_n=0$ and $\sum_{t\ge
  1}t|\rho_t-\rho_{t-1}|<\infty$. Thus, the identity
$\sum_{t=1}^nt(\rho_{t-1}-\rho_t)=\sum_{t=0}^{n-1}\rho_t-n\rho_n$ for
$n\ge 1$ tells us that $\lim_{n\uparrow\infty}n|\rho_n|$ exists, and
the fact that $\sum_{t\ge 0}|\rho_t|<\infty$ implies
$\lim_{n\uparrow\infty}n|\rho_n|=0$. At this point, the same identity
$\sum_{t=1}^nt(\rho_{t-1}-\rho_t)=\sum_{t=0}^{n-1}\rho_t-n\rho_n$ for
$n\ge 1$ leads to $\sum_{t\ge 1}t(\rho_{t-1}-\rho_t)=\sum_{t\ge
  0}\rho_t$, which proves that $\sum_{s\ge
  1}s^2p(s)=\mu+2\mu^3\sum_{t\ge 0}\rho_t$.

\end{document}